\documentclass[letterpaper,twocolumn,10pt]{article}
\usepackage{usenix2019_v3}
\usepackage[utf8]{inputenc}
\usepackage[utf8]{inputenc}
\usepackage{appendix}
\usepackage{filecontents}
\usepackage{cite}
\usepackage{subcaption}
\usepackage{amsmath,amssymb,amsfonts}
\usepackage{algorithmic}
\usepackage{graphicx}
\usepackage{textcomp}
\usepackage[colorinlistoftodos]{todonotes}
\usepackage{bmpsize}
\usepackage{xspace}
\usepackage{lipsum}
\usepackage{pgfplots}
\pgfplotsset{compat=1.14}
\usepackage[official]{eurosym}
\usepackage{versions}
\excludeversion{RedundantContent}   
\excludeversion{Rebuttal}
\excludeversion{OldVersion}
\excludeversion{MyNotes}
\usepackage{multirow}
\usepackage{booktabs}
\usepackage{enumitem}

\usepackage{subcaption}
\usepackage{float}

\usepackage{tcolorbox}
\usepackage{framed}
\usepackage{enumitem}
	
\usepackage{color, colortbl}
\usepackage{xcolor}
\definecolor{LightCyan}{rgb}{0.88,1,1}
\definecolor{apricot}{rgb}{0.98, 0.81, 0.69}
\definecolor{brilliantlavender}{rgb}{0.96, 0.73, 1.0}
\definecolor{babyblue}{rgb}{0.54, 0.81, 0.94}
\definecolor{coquelicot}{rgb}{1.0, 0.22, 0.0}
\definecolor{babypink}{rgb}{0.96, 0.76, 0.76}
\definecolor{bananamania}{rgb}{0.98, 0.91, 0.71}
\definecolor{ashgrey}{rgb}{0.7, 0.75, 0.71}
\definecolor{yellow(ryb)}{rgb}{1.0, 1.0, 0.2}
\definecolor{yellow-green}{rgb}{0.6, 0.8, 0.2}
\definecolor{lightmauve}{rgb}{0.86, 0.82, 1.0}
\definecolor{lightkhaki}{rgb}{0.94, 0.9, 0.55}
\definecolor{lightblue}{rgb}{0.68, 0.85, 0.9}
\definecolor{lightgreen}{rgb}{0.56, 0.93, 0.56}
\definecolor{ashgrey}{rgb}{0.7, 0.75, 0.71}
\definecolor{mediumspringbud}{rgb}{0.79, 0.86, 0.54}
\definecolor{darkseagreen}{rgb}{0.56, 0.74, 0.56}
\definecolor{celadon}{rgb}{0.67, 0.88, 0.69}
\definecolor{teagreen}{rgb}{0.82, 0.94, 0.75}
\definecolor{vividtangerine}{rgb}{1.0, 0.63, 0.54}
\definecolor{peach-orange}{rgb}{1.0, 0.8, 0.6}
\definecolor{peach-yellow}{rgb}{0.98, 0.87, 0.68}

\usepackage{xcolor}

\begin{document}

\date{}

\title{
``Un-Equal Online Safety?"
A Gender Analysis of Security \\
and Privacy Protection Advice and Behaviour Patterns}

\author{
{\rm }
Kovila P.L. Coopamootoo\\
King's College London, UK\\
kovila.coopamootoo@kcl.ac.uk
\and
{\rm Magdalene Ng}\\
University of Westminster, UK\\
m.ng1@westminster.ac.uk
}

\maketitle




\begin{abstract}
There are indications in literature that women do not engage with security and privacy (SP) technologies, meant to keep them safe online, 
in the same way as men do.
To better understand this \emph{gender gap},  
we conduct an online survey with N=604 U.K. participants, 
to elicit SP advice source preference and usage of SP methods and technologies.
We find evidence of un-equal SP access and participation. 
In particular, advice from intimate and social connections (ISC) is more prevalent among women, while online content is preferred by men.
ISC do not closely associate with nor predict the use of SP technologies, whereas online sources (such as online forums, reviews, specialist pages and technology adverts) and training do.
Men are also more likely to use multiple advice sources, that enhances the likelihood of using SP technologies.
Women are motivated to approach ISC due to their perceptions of the advisor (such as IT related expertise, experience and trustworthiness) 
while men approach ISC 
to evaluate options and seek reassurance for their own practices.
This research   
raises questions about the equity of online safety opportunities and makes recommendations. 

\begin{RedundantContent}
While there are a number of security and privacy (SP) information sources, individuals have different preferences for protective SP advice, such as online content or social connections. 
While SoC advice in particular is easily accessible and decreases the SP burden on the recipient, 
it runs the risk of disseminating misinformation, of reducing digital agency and self-efficacy, and may impede improvements in protective SP skills. 
We run a survey with N=604 UK participants to elicit the individuals preferred SP source and usage of SP methods and technologies.
We find SoC advice to be more prevalent among women, and online content among men.
We model how advice source predicts the use of SP technologies.
We synthesise users' rationale for using SoC and the different advice they receive. 
We then run a second survey with N=500 SoC-providing UK participants .....

We observe that protective SoC advice is more prevalent among women than men, and is one of the most used advice sources overall (after general research). 
We find that SoC do not closely associate with nor predict the use of SP technologies, whereas online sources (such as online forums, reviews, specialist pages and technology adverts) and training do.
Our participants report to approach SoC due to their IT related expertise, although this may not constitute specific SP expertise or being equipped to provide SP care and support. 
We also find that the advice received from SoC range from ... to ...

By gaining insight into the prevalence of SoC in providing protective SP advice and individuals' rationale and perception of their SoC, we identify mis-conceptions / generalisations wrt to SoC skills, and expose the quality of advice shared. We show via these qualitative inputs and a quantitative analysis / association of sources with SP technology use that although SoC is prevalent (and the main/only source of advice) for user groups, they impact the content, quality of advice people receive, whether they are effectively protected, as well as whether they are empowered to do so in the future.

Our findings motivates a case for supporting individuals who provide SP support to others, by demonstrating their prevalence, their impact on use of protective SP technologies, and the assumptions recipients make. 
Our findings motivates the further development of a suite of interventions and technology interactions to support the SoC advisors, as well as SP technologies that enable the SoC-supported SP. It also highlights the need to enable users to check their assumptions vis-à-vis their SoC.

Our findings raise questions about the gender inclusivity of SP advice and technologies.

\end{RedundantContent}
\end{abstract}

\section{Introduction}

Digital technologies are a powerful driver of gender equality with the potential to give women and girls access to information, opportunities and resources. But the gender divide persists worldwide, often because of social and gender norms and deep-rooted gender stereotypes, resulting in gendered use of technology~\cite{barbieri2020gender}.
However, while the digital divide with respect to gender is said to be decreasing in developed countries~\cite{van2020digital}, literature offers (only) a handful of empirical research that 
demonstrated that women are poorer in engaging with protective security and privacy (SP) technology although they report more concern~\cite{oomen2008privacy, coopamootoo2022feel,park2015men} compared to men, such as in using firewalls and spamfilters~\cite{oomen2008privacy}, tracking protection~\cite{coopamootoo2022feel}, or engaging with technical privacy protection~\cite{park2015men}.
In parallel, there are reports that women are more at risk of online harms than men~\cite{ofcom2022report}, and 
calls by both international and local organisations for action to ensure women's safety online, such as 
the United Nation Development Fund (UNDP), which 
advocates that it is not enough for women and girls to simply have access to technology and digital skills -- ``\emph{they must also become active agents of change to create a safer and equitable digital future for all}" 
and have called for a safe, affordable, and inclusive Internet~\cite{UNDP2022}. 
In the U.K., the regulator of communications, Ofcom, is also urging technology firms to take actions to keep women safe online~\cite{ofcom2022urgetechfirms}.

So far there has not been much usable SP research effort specifically dedicated to understanding the gender discrepancy, albeit~\cite{oomen2008privacy, coopamootoo2022feel,park2015men}.
This paper reports on research providing insights into the characteristics (what) and causes (why) of this gender discrepancy, thereby demonstrating how technology stereotypes~\cite{adam2005gender} and SP stereotypes~\cite{wei2023skilled} play out in practice.
The first part of our investigation 
focuses on SP access and use
via the research questions:
\begin{itemize}[noitemsep,topsep=0pt]
\item \noindent \textbf{RQ1}: What advice source do individuals use for SP protection, given their gender differences?
\item \textbf{RQ2}: What SP technologies and methods do women versus men use? 
\item \textbf{RQ3}: How does advice source associate with and impact SP usage, given gender differences?
\end{itemize}
In the second part, we look into why individuals seek protective SP advice from intimate and social connections (ISC), referring to family / partners, friends, colleagues and social acquaintances 
and what kind of advice they receive via: 
\begin{itemize}[noitemsep,topsep=0pt]
\item \textbf{RQ4}: For what reasons do women versus men approach ISC for protective SP advice?
\item \textbf{RQ5}: What type of advice do women versus men receive from ISC?
\end{itemize}
We focus on ISC advice because while there are indications that advice from ISC may not cover how to use protective SP~\cite{wu2022sok,murthy2021individually}, ISC remains an accessible and popular (and therefore valuable) source~\cite{wu2022sok} and previous research has already looked into the quality SP advice from the web~\cite{redmiles2020comprehensive}. 

\begin{Rebuttal}
\textcolor{purple}{We stress further that it is not our aim throughout this paper to make judgements about the value of protective advice from ISC, which is not inherently bad.}

Revision Request: 4.	Provide an explanation early on that the source of intimate and social connections (ISC) is not inherently a bad/risky source of advice.

Rebuttal: (R#470B, R#470E): We will make clear in the introduction and discussion that ISCs are not inherently bad. We do not make assumptions about the value of using technology and will ensure the language throughout does not give this impression.
\end{Rebuttal}


\noindent \colorbox{mediumspringbud}{\textbf{Contributions.}}
The main distinguishing contribution of this work, vis-à-vis previous research with regards to SP advice~\cite{redmiles2020comprehensive,rader2015identifying,redmiles2017digital} or socially supported SP~\cite{wu2022sok,murthy2021individually,nicholson2020cyberguardians}, is our relatively large-scale binary gender focus in SP protection access, in particular from ISC, and its implications on SP usage, thereby complementing and extending the few gender analysis in SP research~\cite{oomen2008privacy, coopamootoo2022feel, park2015men, wei2023skilled} who have also looked at women versus men, but not addressed advice source. 
Overall this research (1) provides evidence that women and men have diverging access and participation patterns with SP - which can contribute to the in-equity of online safety and supports arguments of women being more at risk online compared to men, 
(2) supports local and international calls for action to keep women safe online, 
and (3) makes recommendations for multi-stakeholder actions. \\  
\noindent \colorbox{mediumspringbud}{\textbf{Summary of Findings.}}
\noindent \textbf{(1)} \textbf{Women access protective SP differently to men.}
\textbf{(a)} We find evidence that 
women prefer advice from ISC whereas men prefer online content.
\textbf{(b)} Family (only or in combination with another source) is the most reported advice source for women, whereas general research for men.
\textbf{(c)} Women are more likely to report not use any advice source, while men are more likely to report using multiple sources.

\noindent 
\textbf{(2)}
\textbf{Women use SP technologies differently to men.}
Women are more fluent with simple or builtin SP (such as privacy settings, HTTPS, builtin security, security software updates, or passwords)
and non-technology methods for SP protection, compared to men who are fluent with a wider spectrum of SP protection, including more sophisticated methods (such as firewall, VPN, anti-spyware, anti-malware, anti-tracking or multiple factor authentication). 

\noindent 
\textbf{(3)}
\textbf{Online advice and the number of advice sources influence the use of SP technologies.}
\textbf{(a)} 
A preference for advice from ISC does not predict the use of SP technologies, whereas online advice shows 3 to 11 times enhanced likelihood of using SP technologies.
\textbf{(b)} 
An increase in the number of advice sources used, from 1 to 3, gradually increases the likelihood of using SP technologies.

\noindent 
\textbf{(4)}
\textbf{Different motivation for ISC advice.}
Women are 3 times more likely to approach ISC for SP protection, where their motivation (perceived expertise of advisor, experience and trustworthiness) suggest reliance on ISC, while men approach ISC to evaluate options and seek reassurance for their own practices.

\noindent 
\textbf{(5)}
\textbf{Different themes of SP advice.}
A higher \% of women receive authentication advice, whereas a higher \% of men receive malware, fraud and communication / network privacy advice.

\noindent \colorbox{mediumspringbud}{\textbf{Note.}} This paper does not make value judgements on sources of SP advice.
And ISC advice is not inherently poorer than online advice.

\begin{RedundantContent}
\emph{First}, this research builds on existing investigations of SP advice sources~\cite{redmiles2016think}, quality / type of advice~\cite{redmiles2020comprehensive}, SoC supported SP~\cite{murthy2021individually} and motivates further research and technology development for social / collective SP~\cite{das2016social}. 
\emph{Second}, this research evidences a gender disparity, which questions the equality (of access and outcome) of SP advice and protection technologies, with implications to equity of participation in the digital society.
We invetsigate the following research questions and summarise the overall scope of the paper in Figure~\ref{fig:paper_plan}:
\textbf{RQ1} ``\emph{\textbf{what advice source do individuals use and how prevalent is the use of social connections?}}" and 
\textbf{RQ2} ``\emph{\textbf{why do individuals use SoC as advice source?}}".
RQ1 provides a comparison with predominantly US based previous research, while RQ2 provides insights into the motivators for SoC supported SP, that can support further development of human-computer interaction features or interventions both for the recipient and provider of advice, and potentially positively enhance the role SoC plays in awareness and usage of SP technologies. \textcolor{purple}{[findings]}

Next we look into the implications of SoC-supported SP, in particular with regards to the advice type provided, as well as the \textcolor{blue}{association} and impact of SoC on SP methods and technology usage, via \textbf{RQ3} ``\emph{\textbf{what type of protective SP advice do individuals receive from SoC?}}", \textbf{RQ4} ``\emph{\textbf{how do advice sources associate with the use of SP technologies and methods?}}"
and 
\textbf{RQ5} ``\emph{\textbf{how do advice sources predict the use of SP technologies?}}". \textcolor{purple}{[findings]: provides a granular view of the effectiveness of advice sources in SP practice, and SoC do not impact the use of SP technologies, while general research, specialist pages, online forums, online reviews and ad/info from companies do.}

Third, we visualise 
what SoC preference across gender mean for the type of advice and SP technologies people use, via \textbf{RQ6} ``\emph{\textbf{how do men versus women differ in SoC advice preference and rationale?}}" and \textbf{RQ7} ``\emph{\textbf{how do advice received from SoC and use of SP technologies differ between men and women}}". \textcolor{purple}{[findings]}


The COVID-19 pandemic has, so far, triggered three national lockdowns in the UK, consequently
sharply diminishing the amount of physical social contact between individuals and within communities, potentially cutting down on social support, including that for SP practice.
For many, online connection has been the only or main way of social connection.
In addition, more services have been technologised~\cite{}, 
inducing individuals to spend more time online, engaged in a larger spread of online activities, thereby also potentially increasing their likelihood of being victimised in cybercrimes~\cite{}.
This is evidenced by reports of the rise in number and variety of online scams during the pandemic~\cite{}.

We conduct research into how social relationships promote and maintain protective behaviour. 
We investigate as research questions:
\end{RedundantContent}



\section{Background}
In this section, we look into literature addressing security and privacy (SP) inequities, review previous work on SP advice, as well as social connections in relation to protective SP.

\subsection{(Gender) Inequity in Security \& Privacy}
\label{sec:gender_background}
While the digital divide broadly conceptualises how societal diversity impart differential technology access, skills and outcomes, that reinforces societal inequalities~\cite{van2020digital},
research across disciplines has touched on inequity with respect to privacy~\cite{buchi2021digital,redmiles2021feelings}, with particular mention to gender~\cite{oomen2008privacy, park2015men, coopamootoo2022feel,allen1988uneasy,mackinnon1989toward}.

First, 
a handful of recent research has conceptually addressed privacy inequality, in particular using the conceptual scholarship of the digital divide to discuss how privacy contributes to deepening or reproducing one's social standing, framing privacy inequity as a cause or consequence of unequal digital participation~\cite{park2021privacy}, and arguing how the unequal distribution of privacy is a societal problem, 
where online privacy is sensitive to social inequalities pertaining to age, education and gender~\cite{buchi2021digital}, and how trusting others with personal data, concerns about negative or exploitative online experiences, and feelings of control over one's data, influence differential web uses~\cite{redmiles2021feelings}.

Second, feminist perspectives have a long standing of questioning whether online privacy is on the same side for women as it is for men~\cite{allen1988uneasy,mackinnon1989toward}. 
In particular, that historically women have had the `wrong kind of privacy' such as isolation and confinement, while 
what they merit morally and politically are `the right kinds of privacy', namely, meaningful opportunities for choice-ful seclusion, intimacies, and legal rights of decision about personal life and health, where this true privacy for women would further equality and entail radical transformation~\cite{allen1988uneasy,allen1989privacy}.

Third, empirical research has highlighted that even though women are more concerned than men about their privacy online, they are less likely to engage with protective technologies~\cite{bartel1999investigation}.
In particular, women are more likely to report using less technical, non-technology and simple means to protect themselves online, compared to men who employ more diverse technology means~\cite{oomen2008privacy, park2015men, coopamootoo2022feel}. 
Self-reported cybersecurity behaviour also differ between the gender of employees in organisations~\cite{anwar2017gender} and in general
individuals hold gender stereotypes with regards to SP, where men are expected to be more engaged with SP topics or to behave in SP-enhancing ways, while women are expected to have poor SP confidence~\cite{wei2023skilled}.

\begin{Rebuttal}
-	(R#470A): As Wei et al [89] report, they look into the biased assumptions and stereotypes people hold about SP behaviours. We look into the advice seeking and usage behaviours themselves. We will add a discussion on how our findings compare and can be explained by the stereotypes identified by [89] such as (1) men’s confidence and their lower likelihood to ask for help, (2) women’s poorer confidence and potential impact of the negative stereotypes. Together, [89] and this current paper show gender gaps across nations: UK sample here of N=604 and US sample N=202+190 in [89]. Both complement each other in being about binary genders, but are different in the research questions themselves.
-	(R#470D): We distinguish from [20, 61, 62] who do not address SP advice source and their link with SP technology engagement: [20] focused on feelings and behaviour with regards to tracking only with a UK/EU sample, [61] looked at how risk perceptions associate with behaviour across pre-defined SP strategies such as anonymous remailers across a Dutch sample, and [62] investigated how privacy behaviour and confidence differ by gender across a US sample.

Say also that we focus on inequity in access and outcome whereas Wei look into the diff assumptions / beliefs / stereotypes between gender.

\end{Rebuttal}

\begin{RedundantContent}
These three groups of research efforts support the case for more \emph{inclusive SP}, a new strand of efforts within usable SP,  which proposes elevating 
consideration of people’s abilities, characteristics, needs, and values as first-class design requirements for security and privacy mechanisms~\cite{wang2018inclusive}, and recommends that research efforts address inclusivity gaps, where
security training and education efforts currently do not cater to individuals with different abilities, characteristics, needs, identities and values~\cite{das2020humans}. 

Using the conceptual framing of the digital divide into levels of access inequity, skills inequity and use inequity~\cite{dimaggio2004digital, red book}, we focus on how women versus men access SP methods and technologies and its implications on differential use. This investigation provides insight into the effectiveness of the SP advice and use ecosystem, across user groups with diverse characteristics.
\begin{figure}[h]
	\centering
	\includegraphics[keepaspectratio, width=.6\columnwidth]{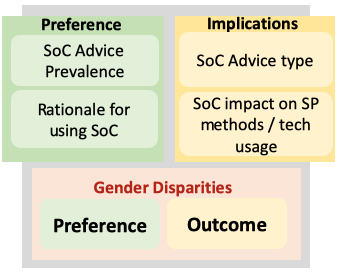} 
	\caption{Overall scope of the paper}
	\label{fig:paper_plan}
\vspace{-0.7cm}	
\end{figure}
\end{RedundantContent}

\subsection{Security \& Privacy Advice Source}

Previous research has evidenced that individuals access security and privacy (SP) advice from various sources, often covering various aspect of SP. 
Advice sources and ways for individuals to learn about SP can be considered as (1) informal while referring to family, friends~\cite{das2014effect}, coworkers~\cite{posey2014bridging}, peers~\cite{nicholson2020cyberguardians}, or news articles; (2) semi-formal referring to webpages from third parties (including online Government webpages, retailers, vendors of software or security-, privacy-focused organisations such as Privacy International)~\cite{rader2015identifying,furnell2014security,james2013determining}; and (3) more formal sources including training and education.

While a high percentage of internet users are not aware of protective SP technologies online~\cite{coopamootoo2020usage}, other SP information such as the source of threats, how they materialise and their consequences may be gained from informal sources.
In particular, personal stories from non-experts (family, friends, peers) have been seen to focus on who the source of threat is, while news articles focus on noteworthy descriptions of incidents and advice relevant to the wider society~\cite{rader2015identifying}.
Of semi-formal sources, webpages are thought to be most authoritative, with their aim to educate Internet users who turn to these when seeking expertise online. They communicate concerns that organisations or governments think non-experts should be aware of, such as Privacy International~\cite{privacyInt2022}, 
the Electronic Frontier Foundation~\cite{eff2022}, 
Microsoft~\cite{microsoft2022} or the U.K.'s National Cyber Security Centre~\cite{NCSC2022}.


In general~\cite{redmiles2016learned}, 
within households~\cite{murthy2021individually}, for particular demographics~\cite{mehrnezhad2022can,murthy2021individually,nicholson2019if,nthala2018informal}, or in technology contexts involving collective ownership of personal information such as social media and smart homes~\cite{das2015role,nthala2018informal}, family, friends or colleagues have been found to be a prevalent source of SP advice. 
Within SP protection contexts such as choosing passwords, authentication, antivirus use, software updating, device /  software prompts have been seen as a prevalent source of advice, followed with online sources, print, TV news or articles 
and online forums~\cite{redmiles2016learned}. 


\subsection{Social connections \& Protective SP advice}
The social dimension of protective behaviour is grounded in the individual's \emph{social capital}. 
Social scientists share the understanding that social capital consists of resources embedded in social relations and social structure, which can be mobilised when individuals wish to increase the likelihood of purposive actions~\cite{lin2002social}. 
In particular, social capital is the total actual or potential resources individuals have access to through their social network~\cite{bourdieu1986forms}, and it includes physical, emotional (instantiated as emotional or social support) and informational (such as advice or novel information) resources~\cite{lin2002social}. 
Social capital has been investigated with regards to SP, in the context of negotiating privacy and information disclosure on social network sites~\cite{stutzman2012privacy,ellison2011negotiating,chen2016empirical}, 
while recent research has conceptualised how social groups (intimate relationships, families and households, social acquaintances and the public) influence others' SP behaviours, in particular via reliance on these social connections for advice and knowledge~\cite{wu2022sok}.

\textbf{\emph{Within households:}} Members of households providing informal support to those 
within the household, have been referred to as \emph{SP stewards} or \emph{self-appointed technology managers}~\cite{das2014effect,murthy2021individually}, who support less technically versed household members (such as older adults) and  
may also establish guidelines for technology usage for the whole family~\cite{murthy2021individually}. However SP stewards are themselves not SP experts and may have gaps in knowledge, and often do not focus on SP technologies (tools and settings) but around past experiences. They may also impose their threat and protection models on household members and control technology use, thereby suggesting paternalism and loss of digital agency.
For younger age groups, parents have been seen to provide advice and help to their children~\cite{kumar2017no}.
Overall, it is thought that those with lower internet skills take advice from family and friends 
and engage in fewer SP practices, thereby potentially increasing the vulnerability of these disadvantaged users~\cite{redmiles2016learned}.

\textbf{\emph{Outside the home:}} Older adults are thought to seek advice from peers such as \emph{cyberguardians}, who are members of a community providing peer-to-peer SP support~\cite{nicholson2020cyberguardians}, and who 
prioritise the availability of a provider of information~\cite{nicholson2019if,nthala2018informal}. 
Protective SP information from social circles has also been found in notifying others about what has been experienced~\cite{das2014effect}, learning lessons from others' shared stories~\cite{rader2012stories}, seeking out information in response to security incident~\cite{redmiles2019should}, or being influenced by others' security feature adoption such as on social media~\cite{das2015role}.
In addition, those with higher internet skills and socio-economic status are thought to take advice from their workplace and have the technical skills to learn from experience~\cite{redmiles2016learned}.

While overall, user characteristics including age~\cite{nicholson2020cyberguardians,murthy2021individually}, education level~\cite{redmiles2017digital}, skills and 
socio-economic status~\cite{redmiles2016learned} 
impact from whom individuals take SP advice, and the advice source in turn impacts the type of advice individuals receive and consequently their experience of SP~\cite{redmiles2017digital}, 
to our knowledge, previous research efforts have not focused on how women versus men receive, seek or gain SP advice and how such behaviour impact their protective behaviour.

\subsection{Research Gap \& Contributions}
First, this paper expands the third category of research described above in Section~\ref{sec:gender_background}, empirically extending knowledge on the previously identified `gender gap'~\cite{oomen2008privacy, park2015men, coopamootoo2022feel}, but
differs from them with a focus into the advice seeking and broad usage behaviours named by participants themselves within a U.K. sample. In comparison, Wei et al. looked into the biased assumptions and stereotypes people hold about SP behaviours with a U.S. sample~\cite{wei2023skilled}, Coopamootoo et al. focused on feelings and behaviour with regards to online tracking only with a U.K. and European sample~\cite{coopamootoo2022feel}, Oomen \& Leenes looked at how risk perceptions associate with behaviour across pre-defined SP strategies such as anonymous remailers within a Dutch sample~\cite{oomen2008privacy}, and Park investigated how privacy behaviour and confidence differ by gender across a U.S. sample~\cite{park2015men}.
In addition, we look into inequity in both access and outcome, thereby also contributing to the digital divide literature~\cite{van2020digital}.

Second, this research also complements the growing usable SP literature investigating SP advice source~\cite{redmiles2020comprehensive} and social SP~\cite{wu2022sok}, in particular research from 
(1) Rader et al. about diverse type of SP advice across sources~\cite{rader2015identifying}, by providing a granular view of how SP advice source choices link to protection; 
(2) Redmiles et al. about the impact of socio-demographics on advice source preference and the prevalence of a security divide across skills level~\cite{redmiles2017digital}, 
and Geeng et al. about the barriers to the effectiveness SP advice for the LGBTQ+ community~\cite{geeng2022like},
by providing a deep binary gender analysis;  
(3) Redmiles et al.'s suggestion that individuals may feel that ISC are experts~\cite{redmiles2016learned},
by providing a granular view of the reasoning for choosing ISC advice versus not; and
(4) Murthy et al.'s~\cite{murthy2021individually} and Nicholson et al.'s~\cite{nicholson2020cyberguardians} qualitative investigation on how social connections support SP protection, by providing a larger sample study with mixed methodology.

\begin{RedundantContent}
\subsection{Social Capital} 

The role of social structure in understanding the nature, sources and generally positive effects of social relationships and supports.
3 distinct aspects of social relationships:
(1) social integration - their existence or quantity -------social integration or isolation usually refer to the mere existence and quantity of social relationships.

(2) social networks - formal structure; 
----refer to structures existing in dyadic ties (reciprocity, multiplexity, frequency) or among a set of relationships (e.g. their density, homogeneity and boundedness)

(3) social support - the functional or behavioural content of social relationship - the degree to which they involve flows of affect or emotional concern, instrumental or tangible aid, information and the like.

\textcolor{purple}{The existence of social relationships is a necessary precondition or cause of network structure, and both of these may influence sentiments of social support} -- See Figure 1 in House 1987.\\

\end{RedundantContent}

\begin{RedundantContent}
\section{Aim}
\subsection{How relationships support protection?}
\label{sec:aim-relationship_to_support_protection}
First, we build on previous research 
~\cite{murthy2021individually,nicholson2020cyberguardians,redmiles2017digital,redmiles2016learned}, in investigating  the potential for social relationships to promote and maintain protective behaviour. 
In particular we investigate whether and how individuals seek or receive protective SP information and/or support from their social relations. 

We inquire about the supportive attributes of social relationships and across age groups and gender:
\begin{itemize}
\item (1) existence, quantity, type of social relationships that individuals can rely on for protective support
\item (2) the structure of the network of relationships/social network - size, density, reciprocity
\item (3) the type and quality of support people are offered/receive vs seek from social contacts: information (diy), guidelines/how-to, do-it-for-me, sustained care vs one-time, support always available/when needed vs very competent support
\item (4) perception of usefulness of support for privacy, perception of online safety after support
\end{itemize}

\subsection{Impact of Social Capital}
\label{sec:aim-impact_of_capital}
Second, we investigate how social capital (general social capital and capital dedicated to protection) influences 
\begin{itemize}
\item (1) the characteristics of SP support individuals seek and/or receive (3 \& 4 in first study), 
\item (2) the SP behaviours they individually or collectively employ, as well as 
\item (3) their susceptibility and experience of cybercrime victimisation.
\end{itemize}

\subsection{Impact of social distancing}
Third, while COVID-19 has induced physical social distancing and resulted in individuals spending more time online, 
thereby potentially changing how individuals find protection, their SP practice and whether they are victimised.
We therefore inquire about social support (Section~\ref{sec:aim-relationship_to_support_protection}) and the impact of capital (Section~\ref{sec:aim-impact_of_capital}), (1) in the last year, as well as (2) how these differ to pre-COVID-19 pandemic.
\end{RedundantContent}

\section{Method}
We designed and conducted a user study via an online survey methodology in May 2021.

\subsection{Survey Design}
\label{sec:survey_design}
The first page of the survey consisted of an information section, followed by opt-in consent.
This was followed with five parts of mandatory questions as described further.\\
\noindent \textbf{Part 1:} 
We elicited engagement with SP via an open-ended question ``\emph{What privacy and security methods or tools do you most often use online?}", inviting participants to name three to five of them,
similar to previous research where participants were found to provide between three to five privacy methods when queried~\cite{coopamootoo2020usage}.

\begin{Rebuttal}
R#470B: Explain artificial cap in S&P methods self-report
We asked participants for 3-5 SP methods they `often' use, similar to the method in [18], and not expecting a long list. We will discuss the implications of this cap in limitations.
\end{Rebuttal}

\noindent \textbf{Part 2:} 
We queried participants about the ways they become aware of SP methods and technologies via another open-ended question 
``\emph{How do you usually become aware of, learn about or find technologies/methods for protecting your privacy and security online?}", inviting participants to name all the sources they use. 

\noindent \textbf{Part 3:} 
We then asked participants if they would approach ISC if they needed SP advice, via a Yes/No response, followed with an open-ended question on their reasoning.
The questions were set as ``\emph{At any time, if you need advice/help to protect your privacy and security online, do you usually approach your intimate and social connections (such as family, friends, coworkers) for help?}" and ``\emph{Please explain why you replied Yes or No to the previous question.}" 

\noindent \textbf{Part 4:} 
We further queried into the example and type of advice they had received before, via ``\emph{If you have received advice/support from intimate and social connections, about technologies/methods to protect your privacy/security online, please provide examples of these advice/support.}"

\noindent \textbf{Part 5:} 
The last section of the survey consisted of (1) demographic questions on age, gender and computing/IT background; followed with 
(2) a digital skills questionnaire via Van Deursen et al.'s Internet / digital skills instrument across 5 skill types, as provided in~\cite{van2014measuring}, which consists of 35 items, organised across 5 digital skills (operational, information navigation, social, creative and mobile), administered with a 5-point Likert from `Not at all true of me' to `Very true of me'.
Compared to previous research on the influence of \emph{technical skills} (measured via technical web skills~\cite{hargittai2012succinct}) on advice source preference~\cite{redmiles2016learned},
our choice of the digital skills instrument~\cite{van2014measuring}
follows Helsper's theoretical model of the digital divide~\cite{helsper2013distinct}, with attention to the relationship between digital skills and engagement with ICT, and whose development involved a U.K. sample~\cite{helsper2013distinct,van2014measuring}.
(3) We also administered Franke et al.'s Affinity for Technology Interaction (ATI) scale~\cite{franke2019personal} (to our knowledge novel in the SP context), which consists of 9 items, administered with a 6-point Likert from `completely disagree' to completely agree'.
ATI assesses individuals' tendency to actively engage in intensive technology interaction (that is how they would approach, avoid or cope with new technical systems)~\cite{franke2019personal}, as a key personal resource for coping with technology. In technology interaction, ATI is manifested as a tendency to approach and explore new systems and functions more actively for problem-solving, versus a tendency to avoid interaction with new systems to prevent experiencing problems with technical systems.

We note that Prolific selection criteria was set to women and men only, but the gender survey question was open to `women, men, non-binary or other’. with only 1 participant identified as non-binary (not considered in analysis).

\subsection{Participants}
\label{sec:participants}
We recruited $N=600+$ participants via Prolific Academic's UK sample pool. 
The study lasted between 10 -15 minutes. Participants were compensated at a rate of £7.5 per hour. 

After removing $14$ incomplete responses and $1$ response self-reporting as non-binary, we ended up with a sample of $N=604$ participants, with $89\%$ identifying with white ethnicity.
Our sample was balanced with the binary gender of men and women, with approximately $50\%$ women and $50\%$ men (more specifically $n=303$ women and $n=301$ men), to expand knowledge on the
previously identified gender gap which specifically compared women to men 
~\cite{oomen2008privacy,park2015men,coopamootoo2022feel}. 
We also balanced our sample across age, with approximately $10\%$ of participants across each of 10 age groups from 18 to 65+, because previous research has shown that age impacts where individuals gain SP advice, with a particular influence on whether they gain advice from family and friends~\cite{das2014effect, murthy2021individually, nicholson2019if, nicholson2020cyberguardians}.

The sample had a higher proportion of university graduates (at $51.1\%$) than the UK population (noted at approximately $42\%$~\cite{ONS2021graduates,standoutcv2022graduates}), 
including $38.2\%$ of university degrees at undergraduate level, $11.8\%$ at masters level and $1.5\%$ at doctorate level.
Education level was similar across gender, as shown in Table~\ref{tab_gender-education} in the Appendix, where $51.2\%$ of the women group versus $51.9\%$ of the men group, had a university degree (undergraduate to PhD combined).
This differs from the U.K. population, where women are more likely to go to university than men~\cite{advanceHE2021equality,UKParliament2021equality,bbc2017record}.

Overall, $16.5\%$ of participants (n=100) reported to have an IT / computing background, that is to have education or to work within the field of IT, computer science or computer engineering, which pertained to approximately $10\%$ of the women group and $23\%$ of the men group.
The gender difference in IT / computing background in our sample is much smaller than that in the U.K. population~\cite{stem2022women}.
(Note that in the results reported in Section~\ref{sec:results}, the gender differences observed were the same for the whole N=604 and the n=504 without IT / computing background.)

We also measured participants' digital skills via Van Deursen et al.'s Internet / digital skills instrument~\cite{van2014measuring} (detailed further in Section~\ref{sec:survey_design}) across 5 skill types.
There was no difference in information navigation, social and mobile digital skills, but there was a slight difference in operational (mean difference = 1.7) and creative (mean difference = 3.5) digital skills, between gender. The scale reliability Cronbach $\alpha$ across the 5 digital skills varied between $.824$ to $.910$.
In addition, employing Franke et al.'s Affinity for Technology Interaction (ATI) scale~\cite{franke2019personal} (detailed further in Section~\ref{sec:survey_design}), we observed a significant difference in ATI between our women ($M=3.37$, $SD=.92$) and men ($M=3.86$, $SD=.97$) participants, with men responding with higher ATI ($p<.001$). ATI had a Cronbach $\alpha$ of $.825$.


\subsection{Ethics}
Our study protocol was approved by Newcastle University's Ethics Board before the research commenced. We also followed the ethics guidelines of King's College London, where the first author was based, for the full data analysis and write-up phases.

We sought participants’ opt-in consent for data collection prior to their responding to the questionnaire and did not collect identifying information.  
Participation in the study was voluntary and anonymous and our participants could drop out of it at any stage.

\subsection{Data Analysis}
\label{sec:analysis}
We employed both qualitative and quantitative analyses, that we describe in this section. 


\textbf{Qualitative.}
The free-form responses collected for parts 2 to 4 of the survey were analysed via a process of inductive content analysis~\cite{neuendorf2002content,riff2014analyzing}, where for each question, we (1) read each response, extracted themes and synthesised responses across categories, (2) developed a codebook which was iteratively refined, 
(3) coded all the responses with the help of 2 coders, and (4) computed inter-rater reliability (IRR).
Part 1's elicitation of SP methods and technologies followed a simple identification of the SP method named.

Our coding approach provided evidence of the presence of codes, which does not provide evidence for their absence. We alleviated the potential effects on our findings by specifically asking participants to say whether they do not use or are not aware (that is have no knowledge of having used an SP technology or method), and similarly for advice source, as detailed in Section~\ref{sec:survey_design} and the questionnaire in the Appendix.

\emph{Part 1-SP methods and technology use:}
Overall, we collated participants report of SP usage to a total of $26$ distinct tools and methods.
We categorised the tools and methods according to \emph{technological} methods and \emph{non-technological} methods (summarised in Table~\ref{tab:SP_categories} of the Appendix). 
We define \emph{SP as technologies} that rely on algorithms, or software programming. In other words, this is where the technology itself has the ability to protect one's privacy and security. 
We define \emph{non-technological SP methods} as comprising of human behaviours and strategies. 

\emph{Part 2-Advice Source for Protective SP:}
We find that individuals become aware of, learn about or find technologies/methods to protect their security and privacy via a list of sources, including those named within previous research~\cite{wu2022sok,redmiles2017digital,redmiles2016think,redmiles2016learned,rader2015identifying}.
We identified (i) a family, friends, co-workers and other social connections category, similar to~\cite{wu2022sok,murthy2021individually,nthala2018informal}, that we group under \emph{intimate and social connections} [note that family included partners, parents, children or siblings];
(ii) an online content category (similar to~\cite{rader2015identifying,redmiles2020comprehensive}) and  (iii) an `other methods' category, that included news and training, as summarised in Table~\ref{tab:sources_categories}. IRR Cohen $k=.950$.



\emph{Part 3-Motivation for seeking ISC advice:}
We categorised participants' rationale for choosing ISC advice as
(i) perception of the technology skills and experience of their ISC, (ii) perception of the qualities (such as trustworthiness, availability, helpfulness) of their ISC, (iii) perception their own skills or (iv) other reasons.
We categorised participants' responses for deliberately not choosing ISC advice according to 
(i) their reference to their own skills such as self-reliance and confidence or (ii) other reasons such as a  preference for another source.
This complements previous research suggesting that individuals use SP advice sources based on their own education skills~\cite{redmiles2016learned}, their trust and convenience perceptions of these sources~\cite{redmiles2016learned},  
and the perceived skills of ISC who may not actually be SP experts~\cite{murthy2021individually}.
IRR Cohen $k=.861$.

\emph{Part 4-Example Advice from ISC:}
We categorised participants report of SP advice received from ISC as (i) specific to an aspect of SP or (ii) general.
Specific SP advice referred to an aspect of SP protection, such as authentication, anti-malware, SP of communication, email and others, as described in Table~\ref{tab:advice_received_from_SoC} of the Appendix.
Responses in the general advice category did not mention a specific SP aspect, but rather
included general SP aspects, such as best practices, how to keep safe online, warnings, installations, or to do it for the participant (Figure~\ref{fig:advice_received_bygender}).
IRR Cohen $k=.907$.

\textbf{Quantitative.}
The occurrence of the themes across gender, advice source, and SP use (methods / technologies) were used in visualisations to depict quantitative differences, $\chi^2$ tests and multivariate analysis to depict associations, and logistic regressions to show predictive influence.
We tested regression assumptions, for example computing the collinearity statistics with the list of sources (referring to regression results in Section~\ref{sec: bayesian}), where VIF ranges from 1.02 to 1.14 across sources, signifying poor correlation between sources.
In addition, given the lack of prior estimates of the effect of gender on SP sources, we used the rule of thumb of >500 participants and at least 10 cases per IV.
Further, we note that content analysis has similarly contributed to empirical research~\cite{bos1999content,kolbe1991content,jones1994accounting}, including large scale quantitative research in the area of SP~\cite{coopamootoo2017whyprivacy,coopamootoo2020usage,coopamootoo2022feel,mehrnezhad2022can}.

\subsection{Limitations}
\label{sec:limitations}
\noindent \textbf{Participant characteristics:} 
The sample had a level of digital skill to enable an account on Prolific platform and was more highly educated than the UK population. A different sample may show variation in SP access and engagement.

\noindent \textbf{Data elicitation:} 
This study relies on self-reports, where the particular SP context may be fraught with gender stereotypes about confidence~\cite{wei2023skilled}, resulting in over-reporting of technology engagement by men and under-reporting by women. 
In addition, social desirability bias may have diverging influence between genders, where women admit more to receive advice from ISC, or to more likely remember receiving help from family, thus conforming to help-seeking stereotypes across genders.
The impact of stereotypes on our findings is further looked at within the Discussion Section~\ref{sec:discussion}.

Self-report surveys have nonetheless been a key method of gathering insights into SP experiences over the years, including research on SP advice~\cite{redmiles2016learned} and SP usage~\cite{coopamootoo2020usage}, relevant to this research, and 
widely used to elicit rich insights into human experiences, perceptions or stereotypes in SP research~\cite{redmiles2016learned,redmiles2017digital,coopamootoo2020usage,wei2023skilled}. 

\noindent \textbf{Analysis:} Our coding for the presence of codes may be argued to come with limitations, which we discussed in Section~\ref{sec:analysis}. However we note that our questions were mandatory and had an option for `don't know', `none' or `other'.

\begin{Rebuttal}
Rebuttal: (R#470B, R#470E): We will address potential impacts of gender stereotypes on self-reports and social desirability bias, in limitations and discussion.

Reviewer E: both genders get advice from family equally often, but that women are more likely to admit to it 
(social desirability bias), or that  when trying to remember what they do, women are more likely to remember activities like getting advice from family that conform to common stereotypes about their identify (stereotype bias).

Revision request: 3.	Revise the Limitations section by focusing on the specific experimental and analysis methods that were used in the study as opposed to the general limitations of user research. Please expand the limitations by adding a discussion on the potential impact of gender stereotypes and social desirability on participants' responses.
\end{Rebuttal}

\begin{OldVersion}
which in general comes with limitations such as reliance on participant memory that can impact on reliability~\cite{junger1999self,yu2010reliability} and arguments of generalisability~\cite{redmiles2019well,tang2022well}. Surveying has nonetheless been a key method of gathering insights into SP experiences over the years, including research on SP advice~\cite{redmiles2016learned} and SP usage~\cite{coopamootoo2020usage}, relevant to this research, and 
widely used, to elicit rich insights into human experiences or perceptions in SP research~\cite{redmiles2016learned,redmiles2017digital,coopamootoo2020usage}. 
Moreover, although self-report can induce response bias~\cite{redmiles2018asking}, the systematic design and analysis and large scale sampling, ensures we can still gain valuable insights from the responses.
\end{OldVersion}

\section{Results}
\label{sec:results}
\subsection{Advice Source Preference} 
\label{sec:sources_results}
We investigate RQ1, that is, what advice source(s) do individuals use for SP protection, given their gender differences, via 2 stages: 
(i) we first describe the categories and types of sources named by participants (we refer to responses from women participants as W\# and M\# from men participants);
(ii) we then look into the gender differences in source type preference, number of advice sources used, including the patterns of multi-source usage.
\subsubsection{Description of Advice Sources Used}
We summarise the categories and types of advice source in Table~\ref{tab:sources_categories}, describe them below and provide example responses in Table~\ref{tab:e.g.response_advice} in the Appendix.

\begin{table}
\centering
\caption{Categories and Example Source (N=604)} 
\label{tab:sources_categories}
\footnotesize
\resizebox{.8\columnwidth}{!}{
\begin{tabular}{llr}
\toprule
\rowcolor{ashgrey}
\textbf{Category} & \textbf{Advice Source} & \textbf{\% Participants} \\ 

\rowcolor{apricot}
&family&16.7 \\
\rowcolor{apricot}
Intimate \& Social  /  & friends& 16.1 \\
\rowcolor{apricot}
 Connection&  face-to-face / offline&5.8 \\
\rowcolor{apricot}
& `from work' &4.5 \\
\rowcolor{apricot}
& colleagues &3.6 \\ 

\rowcolor{lightblue}
& General research e.g. Google & 27.0 \\
\rowcolor{lightblue}
& Specialist pages e.g. Techcrunch & 11.1\\
\rowcolor{lightblue}
Online & Online reviews and recommedations & 8.9\\
\rowcolor{lightblue}
content & Tech adverts, shared company info & 8.6\\ 
\rowcolor{lightblue}
 & Social media content e.g. YouTube, Twitter & 7.9\\ 
\rowcolor{lightblue}
 & Online forums e.g. Reddit & 7.6 \\ 

\rowcolor{lightkhaki}
Other & News, TV shows & 10.8\\
\rowcolor{lightkhaki}
methods & Training  & 5.0  \\
\rowcolor{lightkhaki}
& System prompts and settings & 3.1 \\  
\rowcolor{lightkhaki}
& Consumer magazine & 0.8  \\ 

\rowcolor{lightmauve}
None & Don't know & 16.0 \\
\bottomrule
\end{tabular}
}
\end{table}

\noindent \colorbox{apricot}{\textbf{Intimate \& social connection / face-to-face}}
$37.9\%$ participants reported the ISC category of advice source, with family and friends most named, as shown in Table~\ref{tab:sources_categories}.
`Family' include asking or receiving advice from participants' children, partner, sibling or parent.
Friends and colleagues are self-explanatory.
`From work' include participants speaking about finding out about SP methods and technologies from work, referring to 
their employers, and workplace practices / recommendations,
and `face-to-face/offline' included participants supported by a known contact outside of their family, friends or work realms. 

\begin{OldVersion}
\emph{\textbf{Family}}:
The $16.7\%$ participants who were supported by family included asking or receiving advice from family members, specifically from their children, partner, sibling or parent. Responses included:  
M12 ``\emph{my sons tell me} and i use very few of them - usually long after they have come out";
F38 ``\emph{through my husband}...it is his job";
F63 `\emph{my partner} is a bit of a computer buff and he advises me and keeps my system up to date"; 
F176 ``\emph{I ask my brother} as he works in that line of work".
\emph{\textbf{Friends}}:
$16.1\%$ participants reported to receive advice from friends, often motivated by a perception of the friends' background, such as 
F4 ``\emph{... I also listen to my security-paranoid friends...}";
M18 ``\emph{recommendations from friends who are more tech savvy}".

\emph{\textbf{Colleagues / from work}}:
$3.6\%$ participants gained advice from colleagues, including
F65 ``I read articles shared by colleagues that work in cyber security, and make sure i follow the same security companies they do";
while 
$4.5\%$ participants spoke about finding out about SP methods and tools from work 
not from colleagues per se but from their employers, and workplace practices / recommendations, such as:
M9 ``from my employers IT department";
F282 ``I usually find out from recommendations at work". 
\emph{\textbf{Other known (or face-2-face / offline) contact}}:
$5.8\%$ were supported by a known contact outside of their family, friends or work realms, 
such as
F23 ``i only hear of them from the man who fixes my laptop";
M1 ``local computer store",
M333 ``I ask a young person, preferably an IT student...They tell me what to download, or do it for me".
\end{OldVersion}

\noindent \colorbox{lightblue}{\textbf{Online Content}}
$53.6\%$ participants reported to gain protective SP advice from online content, including general Internet searches, by accessing specialist pages, through online reviews and expert recommendations, advice in online forums and content shared on social media. 

\setlist[itemize]{leftmargin=*}
\begin{itemize}[noitemsep,topsep=0pt]
\item General research: (a) using an Internet search (Google) or (b) starting with a general search that then leads to other online contents offering a comparison or review of protective methods, or (c) explained that they research for protective SP when perceiving a threat. 
\item Specialist pages: reports of using technology webpages and blogs, or webpages providing targeted SP protection advice.
\item Online reviews or recommendations: mentions of expert reviews/recommendations.
\item Technology adverts and shared company information: (a) technology adverts, 
(b) targeted emails / information sent by technology companies or service providers, 
(c) reputable brands or retailers. 
\item Social media content: SP advice shared on social media generally, or shared informally by people they may not personally know or more formally by an organisation such as the police.
\item Online forums: general mentions of online forums, technology related forums or Reddit in particular.
\end{itemize}

\begin{OldVersion}
Participants also referred to reading articles generally online. 
Example responses include: 
M46 ``I usually find out by either googling them online myself..."; 
M170 ``I search the internet for the latest most reliable systems";
M249 ``researching but don't look unless there is a specific threat". 

\emph{\textbf{Specialist pages:}}
$11.1\%$ participants reported to use technology webpages and blogs, or 
webpages providing targeted SP protection advice such as 
M10 ``IT magazine sites \& blogs";
M122 ``I subscribed to a number of websites that are related to technology";
M330 ``they are discussed on tech blogs and forums that I read";
M48 ``I learn from Tech Crunch, Buzzfeed and general news media reportage ...";
M366 ``bleepingcomputer.com"; 
M479 ``magazine PC Pro". 

\emph{\textbf{Online reviews and recommendations:}}
$8.9\%$ participants reported to seeking protective SP advice from online reviews or recommendations from experts such as:
M54 ``I check reviews for online protection and stay with same company if all has gone well in the past".

\emph{\textbf{Technology adverts and shared company information:}}
$8.6\%$ participants reported learning about protective SP via (a) technology adverts (e.g: M132 ``I usually see adverts online about protecting privacy..."; 
M151 ``Advertising from major providers"), 
(b) targeted emails / information sent by technology companies or service providers (e.g. M60 ``...Also get safe browsing info and vpn info sent to my inbox from email companies I use such as ProtonMail"; 
(c) reputable brands or retailers (e.g. F160 ``from microsoft [sic] and other companies themselves"; M172 ``I tend to use well known brands such as Norton"); 
or from 
(d) organisations such as banks (e.g. F296 ``it's provided by my bank").

\emph{\textbf{Social media content:}}
$7.9\%$ participants referred to content shared on social media generally, or shared informally by people they may not personally know or more formally by an organisation such as the police,  with responses including:
F65 ``... eg on linked in [sic], they share security related articles and i check this at least weekly";
M229 ``I usually hear about them via twitter networks..." 
F362 ``...the Police and other agencies post information on social media which I follow";
M449 ``I am subscribed to tech youtubers, who ways to protect". 

\emph{\textbf{Online forums:}}
$7.6\%$ participants named online forums in general, technology related forums or Reddit in particular, including:
M578 ``...Various online communities and forums dedicated to technology and computer use";
M372 ``as member of a number of closed discussion forums topics...where individuals share knowledge and experience as well recommend products"; or
F604 ``I usually hear about it on reddit". 
\end{OldVersion}

\noindent \colorbox{lightkhaki}{\textbf{Other methods}}
$18.8\%$ participants reported finding protective SP advice via other media such as the news, shows on TV, training, prompts via their software or device, or from consumer magazine.

\begin{itemize}[noitemsep,topsep=0pt]
\item News \& TV shows: refer to newspapers, general or technology focused online / TV news, and TV documentaries or programmes.
\item Training: (a) attending a course or training at school or as part of their work organisational requirement or (b) have a computing / SP background, (c) from their own (IT/SP related) role at work thereby implying training.
\item System prompts and settings: (a) to `wait' to be prompted / be advised by the software / device they use or (b) just look into their privacy/security settings.
\item Consumer magazine: non-IT magazines targeting consumer such as `Which?' in the U.K.
\end{itemize}

\begin{OldVersion}
\emph{\textbf{News \& TV shows:}}
$10.8\%$ participants named newspapers, general or technology focused online / TV news, and TV documentaries or programmes, such as
F57 ``occasionally from articles I read in the newspaper / online";
F497 ``by seeing articles about them online, especially on the BBC website";
M459 ``...I learn about the subject from national media, tv \& newspaper articles";
M180 ``watching the News";
F16 ``usually via reliable sites such as BBC or Which or consumer rights programmes (Martin Lewis etc)";
F99 ``tv programmes";
M145 ``I normally get information from programs like gadget show";
F318 ``documentaries on e-safety". 

\emph{\textbf{Training:}}
$5.0\%$ participants found out about SP methods and tools via (1) attending a course or training at school or as part of their work organisational requirement (e.g. M394 ``... an A level course in computers systems and networking"; F30 ``... we have training at work to help us identify and protect ourselves from threats to our internet security"); 
(2) have a computing / SP background (e.g. M405 ``
I have been bought up learning about the need for protection information online"), or (3) from their own (IT/SP related) role at work thereby implying training (e.g. M239 ``I research software engineering and cyber security, it's part of my job").  

\emph{\textbf{System prompts and settings:}}
$3.1\%$ participants reported to not search for SP but (a) to `wait' to be prompted / be advised by the software / device they use or (b) just look into their privacy/security settings, such as
F39 ``I own apple devices and I rely on their regular updates for updating privacy protection on my electronic devices. I don't read details of what protection this provides";
M439 ``I use the settings and FAQ in specific apps or software";
F412 ``...Other times I will check out privacy/security settings to see what I need to do".


\emph{\textbf{Consumer magazine:}}
A very small number, $0.8\%$ participants, reported to find SP tools and methods from consumer magazine, such as 
F24 ``Which? magazine is a good source". 
\end{OldVersion}

\noindent \colorbox{lightmauve}{\textbf{None}:}
Responses were categorised as `none' when participants claimed to (a) not be aware of SP tools and methods or to not look for any advice; and (b) to not look for SP but to use what is perceived as already integrated in their device.

\begin{RedundantContent}
Responses included:
F80 ``I am not usually aware";
F42 ``I become aware usually through the news about breeches of privacy and hacking etc. I do not become aware of technologies/methods for protection and security";
M104 ``I am lazy and rarely check for security methods. I just rely on what is running on my laptop";
F110 ``I don't need to do this; my laptop is a work laptop and is fully protected";
the standard programmes that come with my devices";
M127 ``I use Chromebooks and Linux software so I feel protected enough to not bother looking for any further technologies or methods".
\end{RedundantContent}

\subsubsection{Gender Patterns \& Differences}
We \emph{first} visualise participants' overall responses about preference for SP advice sources across gender, \emph{second} we compute a binomial logistic regression depicting gender differences, and \emph{third} we compare multi-advice usage across gender. 


\noindent \colorbox{mediumspringbud}{Visualisation of Gender Patterns.}
Figure~\ref{fig:qual_sources} (supported by Table~\ref{tab:sources_by_gender} of the Appendix) shows clear patterns of differences between women and men, namely that 
\begin{itemize}[noitemsep,topsep=0pt]
\item a higher proportion of those reporting sources in the ISC category, that is, their family, friends, offline advice (such as a computing support person they know), colleagues or from work practices, are women compared to men;
\item a higher proportion of those reporting to find out about SP methods and technologies via the online content category, such as general research, specialist pages, online forums and reviews, technology adverts or shared online contents, are men compared to women; and
\item among those reporting advice from friends, approximately the same proportion are men and women.
\end{itemize}

In interpreting Figure~\ref{fig:qual_sources}, we note that the proportion depicted is based on the number of participants reporting each advice source (that is the base rate), as shown by Table~\ref{tab:sources_by_gender}, 
where for example that more than twice the number of participants reported to consult family (n=101) than to use shared online content (n=48) or online forums (n=46). 
Although the number of men reporting shared online content versus family, 
is only n=8 higher, we still note that twice the number of men reported shared online content than women and three times the number of women reported advice from family than men.
\begin{figure}
	\centering
	\includegraphics[keepaspectratio, width=1\columnwidth]{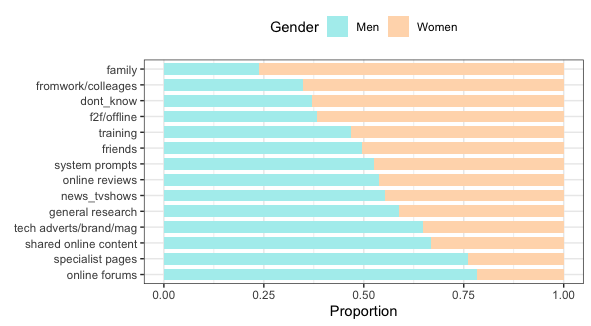} 
	\caption{\emph{participant named sources} across gender}
	\label{fig:qual_sources}
\end{figure}

\noindent \colorbox{mediumspringbud}{Regression Model.}
We compute binary logistic regressions with independent variable (IV) 
women versus men (categorical variable and men as baseline) 
and dependent variable (DV) 
each of the advice sources (categorical variable set at 1 when the advice source is named, and 0 when not named). 
The models with family, work/colleagues, general research, specialist pages, online forums, shared online content and technology adverts are statistically significant (as highlighted in Table~\ref{tab:regression_gender_source}), meaning that the model with gender as predictor fits the data significantly better than the intercept only models.
Compared to men, women are nearly 4X (OR = 3.93, p<.001) and twice (OR = 1.97, p=.029) more likely to report family and work/colleagues respectively as SP advice source.
However, women are also 40\% to 75\% less likely than men to report advice source from the online content category, in particular general research (p=.007), specialist pages (p<.001), online forums (p<.001), shared online content (p=.014) and technology adverts (p=.018) (that is men are between 2 to 4 times more likely to report these advice sources).
These findings loosely corresponds to previous indicative research~\cite{mehrnezhad2022can} and compares with previous research linking lower education with lower likelihood of coworkers or government websites as advice source~\cite{redmiles2017digital}, where our gender groups had comparable education level.


\begin{framed} 
\noindent \textbf{Computing/IT Background.}
For the subsample \emph{without} computing/IT background (n=504),
gender predicts advice source preference similarly as in Table~\ref{tab:regression_gender_source},
and in addition: women are significantly more likely to receive advice from training than men.

\noindent For the subsample \emph{with} computing/IT background (n=100), 
the only significant models involve women 5.9X more likely to receive protective SP advice from family than men, who are 6.9X more likely to receive it from online forums than women.
\end{framed}

\begin{Rebuttal}
\textcolor{purple}{Review Req 10: ... Repeating the analysis in 4.1.2 for technical population...}
Rubuttal: -	(R#470B): We did not compute the model in 4.1.2 for those with technical background only due to the smaller sub-sample size of n=100, but this can be added for consistency.
\end{Rebuttal}


\begin{OldVersion}
\textcolor{blue}{Second, we compute a binomial logistic regression, with women versus men as dependent variables and advice sources as predictors, where each advice source is coded as $1$ when named and $0$ when not named.
The model is significant with $\chi^2 (13) = 91.586$, $p<.001$. 
The model has an accuracy of $66.2\%$ and $R^2$ between $14.1\%$ (Cox \& Snell) and $18.8\%$ (Nagelkerke).
Table~\ref{tab:regression_gender_source} shows that a preference for SP advice from family or from work / colleagues, 
is $73\%$ and $52.4\%$ respectively 
less likely to be from men than women, while a preference for general research, specialist pages, online forums  and technology adverts / brand info / consumer magazines are 2X, 3X, 4X and 2X (respectively) more likely to be from men.
\end{OldVersion}

\begin{table}
\centering
\caption{Regression results depicting the likelihood of women (compared to men) reporting to use each advice source.
Statistically significant models at p<.05 are highlighted. 
OR refers to odds ratio between women and men (the baseline), CI is the confidence interval. and p-value refers to effects of gender.}
\label{tab:regression_gender_source}
\footnotesize
\resizebox{.95\columnwidth}{!}{
\begin{tabular}{l|rrrl}
\toprule
\rowcolor{ashgrey}
\textbf{Source}  & \textbf{OR}  & \textbf{95\% CI} & \multicolumn{2}{c} {\textbf{p-value}}\\
\rowcolor{apricot}\textbf{family}  &\textbf{3.93}& \textbf{[2.41 - 6.42]}&\textbf{<.001}&\textbf{***}\\ 
\rowcolor{apricot} friends  &1.07&	[0.66 - 1.57] &.940 & \\
\rowcolor{apricot}\textbf{from work / colleagues}  &\textbf{1.97} & \textbf{[1.07 - 3.63]} & \textbf{.029}&\textbf{*}\\
\rowcolor{apricot} f2f / offline  & 1.65 & [0.81 - 3.36] &.168&\\
\rowcolor{lightblue}\textbf{general research}  & \textbf{0.61}& \textbf{[0.42 - 0.87]} &\textbf{.007}&\textbf{**}\\
\rowcolor{lightblue}\textbf{specialist pages}  & \textbf{0.27} & \textbf{[0.15 - 0.49]} & \textbf{<.001} & \textbf{***}\\
\rowcolor{lightblue}\textbf{online forums}  &\textbf{0.25} & \textbf{[0.12 - 0.52]} & \textbf{<.001} & \textbf{***} \\
\rowcolor{lightblue}online reviews  &0.84 & [0.48 - 1.48] & .552 & \\
\rowcolor{lightblue}\textbf{shared online content}  & \textbf{0.47} & \textbf{[0.25 - 0.87]} & \textbf{.014}  &\textbf{*}\\
\rowcolor{lightblue}\textbf{tech ad / brand / mag}  & \textbf{0.50} & \textbf{[0.29 - 0.89]} & \textbf{.018}&\textbf{*}  \\ 
\rowcolor{lightkhaki} news / tv shows & 0.78 & [0.46 - 1.30] & .344& \\
\rowcolor{lightkhaki} training &1.14 & [0.55 - 2.39] & .722 & \\
\rowcolor{lightkhaki} system prompts  & 0.89 & [0.36 - 2.22] & .804 &\\
\bottomrule 
\end{tabular}
}
\footnotesize{\emph{Note: Significance codes of $‘***’.001, ‘**’ .01, ‘*’ .05 $}}\\
\end{table}

\noindent \colorbox{mediumspringbud}{Multi Advice Use.}
Table~\ref{tab:atleast1source-by-gender} shows that $15.7\%$ of participants report to not be aware or to not use any SP advice source and nearly $50\%$ report to use only $1$ source while the rest report to use more than $1$ advice source for SP protection. 
A higher $\%$ of women than men report to not be aware or to not use any SP advice source, while a higher $\%$ of men report to use more than $1$ advice source. 


\begin{table} 
\centering
\caption{\% women and men reporting to use zero to multiple advice sources for SP protection} 
\label{tab:atleast1source-by-gender}
\footnotesize
\begin{tabular}{l|rrr}
\toprule
\textbf{\# of advice sources} &\textbf{\% Overall} &  \textbf{\% Women} &  \textbf{\% Men}\\ 
\midrule
Not aware / not use &15.7 & 19.8 &11.6  \\
One source only & 47.7 & 47.9& 47.5 \\
Multiple sources &36.5& 32.3 & 40.8 \\
\bottomrule
\end{tabular}
\vspace{-5mm}
\end{table}

\begin{figure*}
    \centering
    \includegraphics[keepaspectratio,width=.85\textwidth]{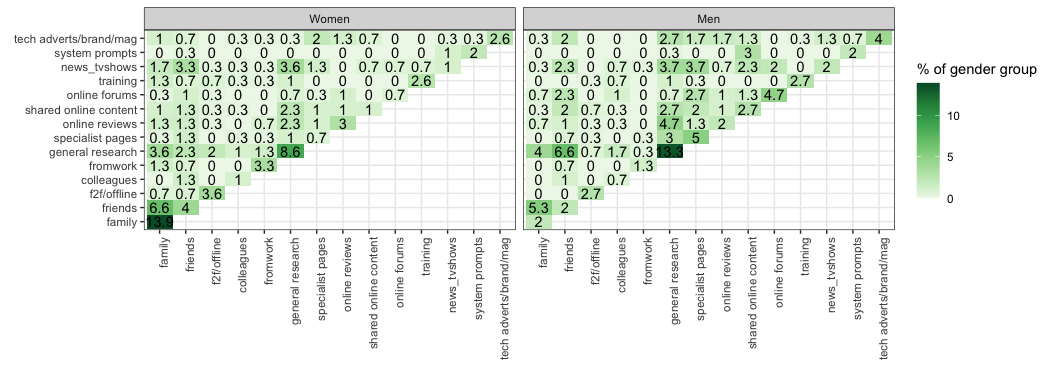} 
    \caption{\% of multiple advice source usage across gender}
    \label{fig:sourcesbysources}
\end{figure*}

We visualise women's versus men's usage of multiple advice sources in Figure~\ref{fig:sourcesbysources}. Again, we notice differences across gender, namely that
(1) a higher $\%$ of women compared to men gain SP advice \textbf{only} from either family (13.9\%), friends (4\%), an offline source (3.6\%), colleagues or work (4.3\%); 
(2) a higher $\%$ of men than women gain advice \textbf{only} from the advice sources from the `online content' or `other methods' categories, except for online reviews - with $13.3\%$ naming general research, $5\%$ specialist pages and $4.7\%$ online forums only;
(3) general research is the most used advice source in combination with another advice source for both men and women; 
(4) general research (only or in combination with another source) is the most used advice source for men, whereas family (only or in combination with another source) is the most used advice source for women, 

\subsection{Security \& Privacy across gender}
\begin{figure}
	\centering
	\includegraphics[keepaspectratio, width=1\columnwidth]{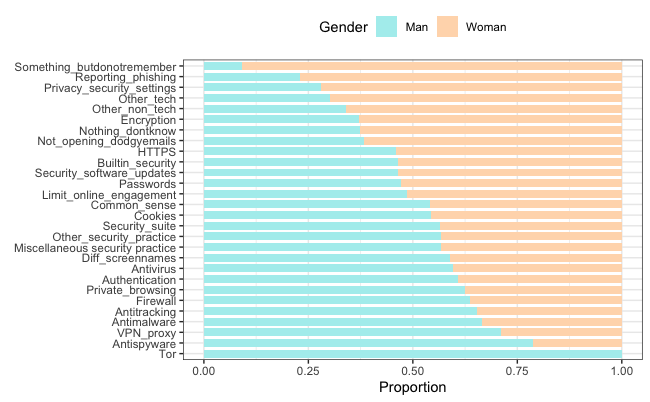} 
	\caption{proportion of women \& men who named particular SP technologies and methods}  
	\label{fig:SP_by_gender}
\end{figure}

We investigate RQ2, that is, what SP technologies and methods do women versus men use. 
We provide the full list of SP technologies and methods named by participants in Table~\ref{tab:SP_categories}, Appendix~\ref{sec:appen_SP_list}, and 
summarise the usage differences between men and women here:
in Figure~\ref{fig:SP_by_gender}, we notice that a higher proportion of men engage with more technological SP, while women have less sophisticated SP behaviour in engaging with builtin type SP and using non-technology methods more than men. 

We find that men are significantly more likely to engage with technology SP, with $\chi^2 (1) = 16.746$, $p<.001$, 
while women are significantly more likely to report not engaging with SP method, with $\chi^2 (1) = 11.461$, $p=.001$.
There is no significant difference across gender for non-technology SP.
This is supported by the contingency Table~\ref{tab:atleast1SP-by-gender}, Appendix~\ref{sec:appen_SP_list}, that summarises women's versus men's engagement with SP methods.

\subsection{Implications of Advice Source Preference}
We investigate RQ3, that is how does advice source associate with and impact SP usage, given their gender differences.
\subsubsection{Association of Advice Source and SP usage}
\begin{figure*}
    \centering
    \includegraphics[keepaspectratio,width=.82\textwidth]{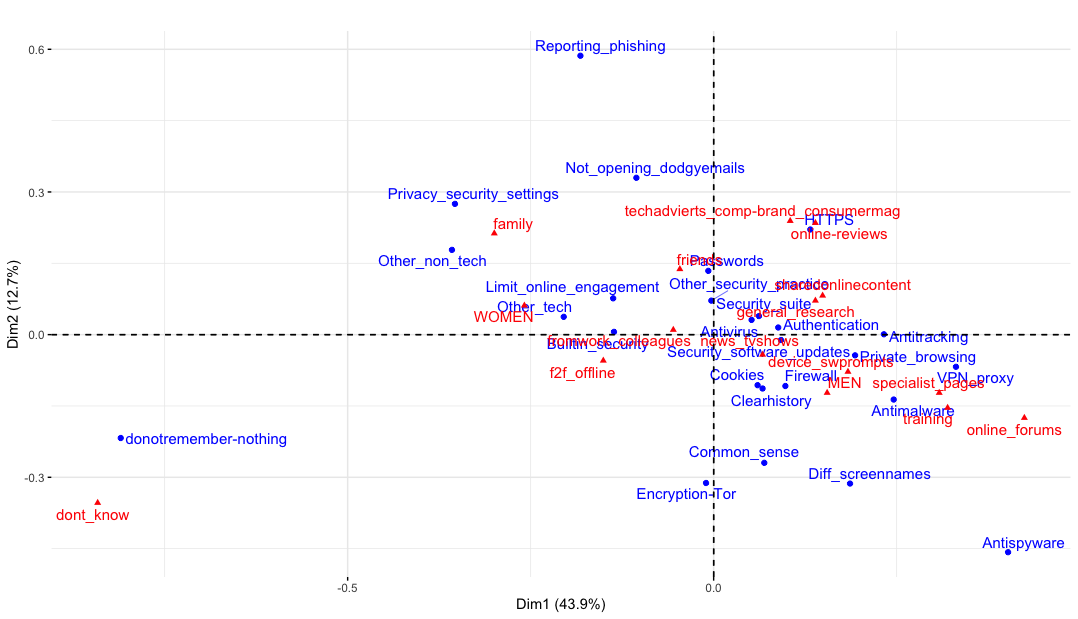} 
    \caption{Spatial Map of Association between SP Technologies \& Methods \emph{(in blue)} and Advice Sources \emph{(in red)}}
    \label{fig:spatial_map}
\end{figure*}

We investigate the statistical association between advice source and use of SP technologies and non-technological methods across gender via 
a Correspondence Analysis (CA)~\cite{greenacre2017correspondence} and depict the strength of association in the spatial map in Figure~\ref{fig:spatial_map}.

We find significant association between SP and advice source, with $\chi^2 = 551.203$, $p<.001$, where
the first dimension (Dim1) accounts for $ 43.92\%$ of the variance in the data while the second dimension (Dim2) accounts for $12.70\%$. Together these two dimensions account for $55.62\%$ of variability. 
The proximity of same colour points demonstrates their similarity in Figure~\ref{fig:spatial_map}, where advice sources (red points) that are closer together on the map have similar SP technologies and methods usage profiles, than those that are further apart, while the proximity of red-blue points demonstrate their association.
Figure~\ref{fig:spatial_map} intuitively shows a range of advice source (red points) with different qualities from (1) not knowing on the left (depicted by `don't know'); to (2) socially connected advice sources (depicted by `family', `friends', `from work and colleagues' and `f2f\_offline' sources; 
to (3) advice sources potentially involving more complex skills (depicted by `training', `specialist pages' and `online forums') on the far right. 
The spatial map is intuitive in showing that the `don't know' end of Dim1 (negative end of x-axis) is more closely associated with not using any SP technologies and method (depicted by `donotremember-nothing').
In comparison, the socially connected advice sources of Dim1 are closer to `other non tech', `other tech', privacy and security settings (usually builtin), other security practice, passwords, builtin security (as well as limit engagement and not opening dodgy emails), and the women gender group. 
`News and TV shows' is more closely associated with firewall, 2FA/MFA authentication. 
Furthermore, the more complex skills end of Dim1 (with training, specialist pages and online forums as advice sources) is closely associated with anti-malware, VPN, anti-tracking or anti-spyware, and the men gender group.

\subsubsection{Impact of Advice Source on use of SP technologies}
\label{sec: bayesian}
We investigate how do advice sources predict the use of SP technologies (1) by the type of advice source and (2) the number of advice sources used.

\noindent \colorbox{mediumspringbud}{\textbf{Advice source type.}}
We choose to focus on SP technologies rather than the combination of technologies and methods, to assess how the different advice sources impact use of technologies in particular.

We aim to compute a binomial logistic regression with the dependent variable `using at least one SP technology' versus `not using any SP technology' and with advice sources as predictors. 
We observe the phenomenon of `quasi-complete separation'~\cite{mansournia2018separation,rainey2016dealing,zorn2005solution} in our data, where the explanatory variable almost perfectly predicts the binary outcome variable~\cite{albert1984existence}. 
In particular, we find that
advice sources such as training, online forums, online reviews, specialist pages and system prompts have \emph{only 1 or 2} participants not using SP technologies, and shared online content has \emph{`less or equal to 5'} participants not using SP technologies as shown in Figure~\ref{fig:quasi_data_separation} of the Appendix, while most participants using these sources reported to using an SP technology.

Separation is a common problem in applied logistic regression with binary predictors~\cite{zorn2005solution}, that can be addressed via a bayesian logit regression~\cite{gelman2008weakly,gelman2013philosophy} with a weakly-informative Cauchy (0, 2.5) prior distribution to regularise the coefficients, as suggested by Gelman et al.~\cite{gelman2008weakly}.
We set up a bayesian logit regression using the \emph{bayesglm} from the \emph{arm} package in \emph{r}~\cite{gelman2013package}.
The model is significant with $R^2=12.4$, AIC 538.29.
Table~\ref{tab:regression} reports that while advice source under the social connections category (such as family, friends, colleagues, from work or another offline contact) do not significantly impact use of SP technologies, advice sources under the `online content' and `other methods' categories significantly predict the use of SP technologies.
For example, those who learn about SP technologies from training, are nearly 10 times more likely to use the SP technology than those who do not learn from training, with $OR=9.62, p=.007$, those who learn about SP technology from online content such as general research, specialist pages, online forums, online reviews are approximately 3X, 8X, 11X, 6X respectively, more likely to use SP technology, compared to those who do not gain advice from these sources.
We also find that individuals benefitting from advice from adverts or targeted information from reputable technology companies, banks or consumer magazines are 5 times more likely to use SP technologies, with $OR=5.21, p=.004$, compared to those who do not gain advice from these sources.

\begin{table}
\centering
\caption{Bayesian logistic regression for using protective SP technology versus not.}
\label{tab:regression}
\footnotesize
\resizebox{.88\columnwidth}{!}{
\begin{tabular}{lrrl}
\toprule
\rowcolor{ashgrey}\textbf{Predictors}& \textbf{OR}  & \multicolumn{2}{c} {\textbf{p-value}}\\
\rowcolor{apricot}family (vs not) &1.10&.734& \\ 
\rowcolor{apricot}friends (vs not) &1.43 &.252& \\
\rowcolor{apricot}work\_colleagues (vs not) & 1.28 & .530&  \\
\rowcolor{apricot}other f2f\_offline(vs not) & 1.17 & .717&\\
\rowcolor{lightblue}\textbf{general research (vs false)}  &\textbf{2.99}& \textbf{<.001}&\textbf{***}\\
\rowcolor{lightblue}\textbf{specialist pages} &\textbf{8.08} & \textbf{ .002} & \textbf{**}\\
\rowcolor{lightblue}\textbf{online forums} & \textbf{11.13}&  \textbf{.004} & \textbf{**} \\
\rowcolor{lightblue}\textbf{online reviews} & \textbf{5.61} &  \textbf{.011} & \textbf{* } \\
\rowcolor{lightblue}shared online content & 2.08 &  .139 & \\
\rowcolor{lightblue}\textbf{ad/info from companies} \& \textbf{consumer mag} &  \textbf{5.21} &  \textbf{.004} & \textbf{**} \\ 
\rowcolor{lightkhaki}news &1.05 &  .909 &\\
\rowcolor{lightkhaki}\textbf{training} & \textbf{9.62} & \textbf{.007} & \textbf{*} \\
\rowcolor{lightkhaki}system prompts and settings &  3.53& .076& \\
\bottomrule
\end{tabular}
}\\
\footnotesize{\emph{Note: Significance codes of $‘***’.001, ‘**’ .01, ‘*’ .05 $}}
\end{table}

\noindent \colorbox{mediumspringbud}{\textbf{Number of advice sources.}}
We compute a binomial logistic regression model 
with predictor variable the number of SP advice sources used and target variable `using at least one SP technology' versus `not using any SP technology'.
The model is significant with $\chi^2 (3) = 57.263$, $p<.001$. It has a good fit ($C=.678$), model accuracy of $80.8\%$ and $R^2$ between $9.0\%$ (Cox \& Snell) and $14.5\%$ (Nagelkerke). 
We find that compared to those not using any SP advice source, those using a single source are 5.4X with p<.001, those using 2 sources are 7.1X with p<.001 and those using at least 3 advice sources are 9.1X with p<.001, more likely to use SP technologies.


\subsection{Motivation for ISC advice}
We investigate RQ4, that is, for what reasons do women and men approach ISC for protective SP advice.
Overall $63\%$ (n=382) of participants responded \emph{Yes} to approaching ISC for protective SP information and advice, while $37\%$ (n=222) responded \emph{No}. 
We \emph{first} describe the rationales provided by participants and 
\emph{second} we look into gender patterns in these rationales.

\subsubsection{Description of Rationales}
We categorise and describe the rationales given for approaching or not approaching ISC below and 
providing example responses 
in Table~\ref{tab:e.g.response_approach_ISC} in the Appendix.

\begin{OldVersion}
\begin{table}[h]
\centering
\caption{Summary of reasons for seeking SP help} 
\label{tab:SoC_help_reasons}
\footnotesize
\resizebox{.87\columnwidth}{!}{
\begin{tabular}{clrr}
\toprule
\rowcolor{ashgrey}\textbf{ISC Help} & \textbf{Category} & \textbf{Reason} & \textbf{\% of N=604} \\ 

\cellcolor{darkseagreen} &\cellcolor{celadon}&\cellcolor{celadon}Technology / security skills&\cellcolor{celadon}13.7 \\
\cellcolor{darkseagreen} &\cellcolor{celadon}ISC & \cellcolor{celadon}Knowledge&\cellcolor{celadon} 13.2 \\
\cellcolor{darkseagreen} &\cellcolor{celadon}Skills& \cellcolor{celadon}Work in IT & \cellcolor{celadon}9.3 \\
\cellcolor{darkseagreen} &\cellcolor{celadon}&\cellcolor{celadon}Experience&\cellcolor{celadon}6.6 \\
\cellcolor{darkseagreen} &\cellcolor{celadon} & \cellcolor{celadon}Up to date & \cellcolor{celadon}1.3\\ \cline{2-4}

\cellcolor{darkseagreen}& \cellcolor{teagreen}& \cellcolor{teagreen}Trustworthy & \cellcolor{teagreen}9.6\\
\cellcolor{darkseagreen}& \cellcolor{teagreen}ISC &\cellcolor{teagreen} Ease of access & \cellcolor{teagreen}2.5\\ 
\cellcolor{darkseagreen}\textbf{YES}& \cellcolor{teagreen}Qualities& \cellcolor{teagreen}Helpful & \cellcolor{teagreen}2.1\\
\cellcolor{darkseagreen}&  \cellcolor{teagreen}& \cellcolor{teagreen}Available & \cellcolor{teagreen}1.5\\
\cellcolor{darkseagreen}& \cellcolor{teagreen}&\cellcolor{teagreen}Reliable &\cellcolor{teagreen} 1.0 \\ \cline{2-4}

\cellcolor{darkseagreen}&\cellcolor{celadon}Own&\cellcolor{celadon}Need help  & \cellcolor{celadon}3.5 \\  
\cellcolor{darkseagreen}& \cellcolor{celadon}Skills&\cellcolor{celadon}Knowledge &\cellcolor{celadon} 1.2\\
\cellcolor{darkseagreen}&\cellcolor{celadon}&\cellcolor{celadon}Confidence  & \cellcolor{celadon}0.5  \\ \cline{2-4}  

\cellcolor{darkseagreen}& \cellcolor{teagreen}& \cellcolor{teagreen}Hear options / mutual &\cellcolor{teagreen}3.3\\ 
\cellcolor{darkseagreen}&\cellcolor{teagreen}Other& \cellcolor{teagreen}Reassurance &\cellcolor{teagreen} 3.0 \\ 
\cellcolor{darkseagreen}&\cellcolor{teagreen}&  \cellcolor{teagreen}Other / none& \cellcolor{teagreen}2.8 \\ 

 \cellcolor{vividtangerine}&\cellcolor{peach-orange}Own& \cellcolor{peach-orange}Self reliance&\cellcolor{peach-orange}8.8 \\
 \cellcolor{vividtangerine}&\cellcolor{peach-orange}Skills& \cellcolor{peach-orange}Better skills & \cellcolor{peach-orange}6.9 \\
 \cellcolor{vividtangerine}\textbf{NO}  & \cellcolor{peach-orange}& \cellcolor{peach-orange}Confidence &\cellcolor{peach-orange}2.3\\ \cline{2-4}

\cellcolor{vividtangerine}  &\cellcolor{peach-yellow}& \cellcolor{peach-yellow}Prefer other source&\cellcolor{peach-yellow}12.7\\ 
\cellcolor{vividtangerine}&\cellcolor{peach-yellow}Other& \cellcolor{peach-yellow}Do not need help /other &\cellcolor{peach-yellow}5.6\\
 \cellcolor{vividtangerine} &\cellcolor{peach-yellow}& \cellcolor{peach-yellow}The one helping others &\cellcolor{peach-yellow}1.3\\ 

\bottomrule
\end{tabular}
}
\end{table}
\end{OldVersion}

\noindent \colorbox{darkseagreen}{\textbf{Yes to approaching ISC for SP.}}
Participants' reasoning for approaching ISC for protective SP is categorised under perception of ISC skills, ISC qualities, participants' own skills, or `other'.

\noindent\textbf{ISC skills.} $39.3\%$ participants referred to perceived ISC skills, most of whom spoke of general knowledge, skills, experience or IT work, and few to SP skills in particular. 

\begin{itemize}[noitemsep,topsep=0pt]
\item \emph{(General) Knowledge of ISC}: (a) knowledge in general terms, (b) expecting their ISC to have more knowledge than them, or (c) somehow pointing to `area of expertise' without being specific of SP.
\item \emph{Technology skills of ISC}: (a) mostly in general, (b) a few named security. 
\item \emph{ISC work in IT}: (a) broad mentions of working in IT / an IT-related field,
(b) pointing to an area of computing (but not security), or (c) very few responses including working in security. [\textbf{(a-b) do not guarantee SP expertise}]
\item \emph{Experience of ISC}: (a) perceived ISC experience in broad terms or (b) ISC would have had experience of identifying the best tools / methods in the past. 
\item \emph{Up to date}: ISC aware of the latest in technology development. 
\end{itemize}

\begin{OldVersion}
\textbf{(1) ISC skills:}
$39.3\%$ of participants referred to at least one reason categorised under skills of their ISC. 

\emph{\textbf{(General) Knowledge of ISC:}}
Among those, $13.2\%$ referred to their SoC's knowledge (1) in general terms, such as F41 ``\emph{Because they are usually very knowledgeable}"; 
(2) that they expect their ISC to have more knowledge than them, such as M114 ``\emph{Usually ask the kids as they know so much more than me}; 
or (3) somehow pointing to `area of expertise' without being specific of SP, such as 
M352 ``\emph{My friends and family tend to know about these things before me so i go them for advice}. 

\emph{\textbf{Technology / Security Knowledge of ISC:}}
$13.7\%$ referred to the (1) \textbf{technology skills of ISC}, with few of them pointing (2) specifically to security.
Example responses naming technology, IT, computer or software knowledge included:
F42 ``\emph{because they \textbf{understand technology} more than me},
M96 ``\emph{as i have family members and co-workers who are \textbf{very tech savvy}}",
M64 ``\emph{I have friends who are \textbf{good with IT}}", 
F94 ``\emph{My daughters \textbf{know more about computers} so I ask them}", 
F188 ``\emph{I ask my husband as he \textbf{knows quite a bit about computers, software} etc}",
M3 ``\emph{Coworkers have vast experience and \textbf{knowledge of cyber security}},
F37 ``\emph{Because I live with my partner who is very \textbf{IT security aware} ...}". 

\emph{\textbf{ISC work in IT:}} $9.3\%$ of participants referred to their ISC working in IT / an IT-related field, 
among whom a couple specifically spoke of ISC working in security. 
Working in IT / IT-related field mentions were (1) broad, 
such as
F28 ``\emph{Husband \textbf{works in IT}}", 
F470 ``\emph{My husband helps me with all this.  He is an IT engineer}",
or (2) pointed to an area of computing, distinct to security, 
such as responses from
M15 ``\emph{My daughter is very diligent and \textbf{her partner is a web developer} so I use their knowledge and advice if needed}",
F29 ``\emph{I have \textbf{friends that are computer programmers} so I know they will tell me what I need to do}",
The very few responses that referred to ISC working in security included
F65 ``\emph{Most of my family and friends and coworkers \textbf{all work in IT Security / Networking jobs}}". 

\emph{\textbf{Experience of ISC:}} $6.6\%$ of participants mentioned (1) ISC's experience in broad terms or (2) that their ISC would have experienced a similar situation or would have identified the best tools / methods. Example responses include: 
M147 ``\emph{They have greater experience than me}",
F72 ``\emph{There's usually someone who's been in that situation before and who doesn't mind sharing what they did}",
F175 ``\emph{someone will likely have come across a similar situation and researched and dealt with it}".
\emph{\textbf{Up to date:}} $1.3\%$ referred to their ISC's awareness of the latest in technology development, such as 
M150 ``\emph{[they] are the most up to date and informed on the subject}". 
\end{OldVersion}

\noindent\textbf{ISC qualities.} $16.2\%$ spoke of perceived qualities of ISC including: 

\begin{itemize}[noitemsep,topsep=0pt]
\item \emph{Trustworthy}: (a) trusting ISC or ISC advice in general or (a) trusting ISC because of their perceived skills and experience.
\item \emph{Easy to access}: ease of access to ISC advice or quick contact.
\item \emph{Helpful}: (a) ISC being helpful or helps them or (b) offering good / helpful advice.
\item \emph{Available}: (a) availability of an ISC they can ask for help or (2) that their ISC is (always) available for such things.
\item \emph{Reliable}: specifically that their ISC is reliable.

\end{itemize}

\begin{OldVersion}
\emph{\textbf{Trustworthy:}}
$9.6\%$ of participants spoke of (1) trusting ISC or ISC advice in general or (2) trusting ISC because of their perceived skills and experience. For example
M137 ``\emph{Because a friend and family member are trustworthy connections}", 
M145 ``\emph{I trust their input and knowledge}", or 
M360 ``\emph{...Some of my friends are very good with technology and understand cyber security far better than me, thus I trust their judgement}",
F560 ``\emph{My father is very experienced with computers and I trust his opinion}". 
\emph{\textbf{Easy to access:}} $2.5\%$ of participants mentioned ease of access to ISC advice or quick contact, 
including 
M68 ``\emph{easy access}",
M112 ``\emph{Because I find it easier to get advice from others}", and
F191 ``\emph{they know quicker than me finding it}".
\emph{\textbf{Helpful:}} $2.3\%$ of participants referred to ISC (1) being helpful or helps them, and (2) offering good / helpful advice.
M484 ``\emph{They are helpful}",
F76 ``\emph{my partner helps me}",
M575 ``\emph{They have helped me improve my online habits tremendously}",
F283 ``\emph{As they generally offer good advice}",
F586 ``\emph{They can provide useful and helpful information}".
\emph{\textbf{Available:}} $1.5\%$ of participants spoke (1) of the availability of a ISC they can ask for help or (2) that their ISC is (always) available for such things, such as
F230 ``\emph{...computer techie fellah.  He's easily available}",
M78 ``\emph{There is usually someone I know I can ask for advice}",
F209 ``\emph{Because my friend is always contactable}".
\emph{\textbf{Reliable:}} $1\%$ specifically said that their SoC is reliable, such as 
M12 ``\emph{they are the most reliable and know more than me}", or
M140 ``\emph{I can rely on my family}".
\end{OldVersion}

\noindent\textbf{Participant's own skills.}
$7.1\%$ of participants referred to their own skill level, in particular that (1) they need help or needing particular advice or information that they cannot find by themselves, (2) they lack SP knowledge or (3) have poor confidence.

\begin{RedundantContent}
\begin{itemize}[noitemsep,topsep=0pt]
\item \emph{Need help}: (a) participants' need for help or support or (b) worry and needing particular advice or information that they cannot find by themselves.
\item \emph{Knowledge}: participants' own lack of lack of knowledge.
\item \emph{Confidence}: participants' poor confidence/
\end{itemize}
\end{RedundantContent}

\begin{OldVersion}
\emph{\textbf{Need help:}} $3.5\%$ of participants referred (1) to their need for help or support, (2) to worry and needing particular advice or information that they cannot find by themselves, such as
F500 ``\emph{yes, i ask for help \textbf{when needed} from family, friends, other people}",
M231 ``\emph{I \textbf{just need support sometimes}}", 
F225 ``\emph{\textbf{If I’m worried} I ask my son what I should do}".
\emph{\textbf{Knowledge:}} $1.2\%$ referred to their own lack of lack of knowledge, such as 
M18 ``\emph{because I'm not that tech savvy and learning about online technologies doesn't really interest me at all}". 
\emph{\textbf{Confidence:}} $0.5\%$ of participants spoke about their poor confidence such as 
F102 ``\emph{Not confident enough to do on my own}".
\end{OldVersion}

\noindent\textbf{Other reasons.}
$10.6\%$ gave reasons that were different from the categories defined above, such as: 

\begin{itemize}[noitemsep,topsep=0pt]
\item \emph{Hear options/mutual:} (a) to find out about the options available, (b) about the mutual sharing of advice and experiences.
\item \emph{Reassurance:} approach ISC (a) for a second opinion, (b) for reassurance or to ensure the correctness of their SP practice, or (c) to avoid making mistakes.
\item{Other/none}: a small number of participants provided miscellaneous statements that do not fall into the reasons named above.
\end{itemize}

While advice from ISC is not necessarily risky advice, unfortunately the heuristics named
by women participants, such as work in IT, trustworthiness, or ease of access (complementing previous research~\cite{redmiles2016think}), do not guarantee ISC actually have SP skills to advise and enable learning, with previous research noting that family are not necessarily SP experts~\cite{murthy2021individually}. 
In contrast the men's rationale indicate a more analytical dialogue (such as evaluating options).

\begin{OldVersion}
\emph{\textbf{Hear options/mutual:}} $3.3\%$ of participants aimed (1) to find out about the options available, and (2) about the mutual sharing of advice and experiences, such as 
M572 ``\emph{
to see what options are out there}",
M152 ``\emph{Me and my friends will always share useful information with each other about anything happening online}", or
F282 ``\emph{Solidarity and advice}". 
\emph{\textbf{Reassurance:}} $3.0\%$ of participants approach SoC (1) for a second opinion, (2) for reassurance or to ensure the correctness of their SP practice, or (3) to avoid making mistakes.
M186 ``\emph{It's better to get a second opinion on things, since you might have missed some important detail}",
M46 ``\emph{I like to get reassurance that what I'm doing is correct and it is the same or similar to how they are doing it}",
F590 ``\emph{Because I want to be sure that I am doing the correct thing}". 
\emph{\textbf{Other / none:}} $2.8\%$ of participants provided miscellaneous statements that do not fall into the reasons named above. 
\end{OldVersion}

\noindent \colorbox{vividtangerine}{\textbf{No to approaching ISC for SP.}}
Participants' reasoning for \textbf{\emph{not}} using their ISC for protective SP information is categorised under perceptions of their skills and other reasons, described below. 

\noindent\textbf{Participant's own skills.}
$17.5\%$ of participants provided reasons referring to their better skills, their self-reliance or their confidence. 

\begin{itemize}[noitemsep,topsep=0pt]
\item \emph{Self reliance}: (a) they prefer to rely on themselves, (b) that they can figure out SP protection on their own.
\item \emph{Better skills}: (a) that their own (technology) skills were more advanced than that of ISC, (b) that ISC know less than them / did not have the skills.
\item \emph{Confidence}: being confident in finding SP information or in their ability.
\end{itemize}

\begin{OldVersion}
\emph{\textbf{Better skills:}} $6.9\%$ participants stated that their own (technology) skills were more advanced than that of ISC or that ISC know less than them / did not have the skills, such as
M2 ``\emph{...because they know less than me}", 
M131 ``\emph{I don't think my friends and family know much about anything privacy related}",
M8 ``\emph{I am the IT expert}",
M32 ``\emph{It's one of my areas of expertise and I know far more about it than most friends, family, and coworkers}".
\emph{\textbf{Confidence:}} $2.3\%$ of participants spoke of being confident in finding SP information or in their ability, such as 
M107 ``\emph{I work in IT. I am confident in finding information myself}" or
M127 ``\emph{I'm confident in my own ability to find what I need}".
\emph{\textbf{Self reliance:}} $8.6\%$ participants mentioned that they prefer to rely on themselves, or that they can figure out SP protection on their own.  
M74 ``\emph{I can rely on my own research}",
M95 ``\emph{I can do it myself}",
M517 ``\emph{I tend to be quite self sufficient and find things out on my own}", or
F555 ``\emph{I can figure things out on my own}".
\end{OldVersion}

\noindent\textbf{Other reasons.}
$19.5\%$ of participants referred to other reasons to not approaching ISC for SP protection, such as (1) preferring other sources (such as online sources), (2) not needing help or other reasons such as not encountered issues (3) that they are the ones helping others.

\begin{OldVersion}
\emph{\textbf{Prefer other source:}} $12.7\%$ of participants said that they preferred other sources, such as  
M1 ``\emph{i prefer government websites for info}",
M35 `\emph{Usually there is a wealth of expert information online so there is no need to consult family and friends - they are rarely the experts!}".
\emph{\textbf{The one helping others:}} 
About $1.3\%$ of participants said that they were the one helping others, such as 
M10 ``\emph{They would expect me to be able to tell them}".
\emph{\textbf{Do not need help / other:}} $5.6\%$ of participants stated that they do not need SoC support or provided other reasons that did not fall into the previous themes, such as
M602 ``\emph{Because I don't feel the need too}",
M490 ``\emph{Never lost data}",
M196 ``\emph{not sure}".
\end{OldVersion}




\subsubsection{Gender Patterns \& Differences}
$75.9\%$ of the n=303 women participants and $50.5\%$ of the n=301 men participants
responded with \emph{Yes} to approaching ISC for protective SP information and advice.
We looked into the association between gender and approaching versus not approaching ISC via 
a $\chi^2$ test. 
We find a significant association between gender and whether individuals usually approach ISC, 
with $\chi^2 (1) = 41.939$, $p<.001$, Cramer $V=.264$. 

\begin{figure}[h]
	\centering
	\includegraphics[keepaspectratio, width=.9\columnwidth]{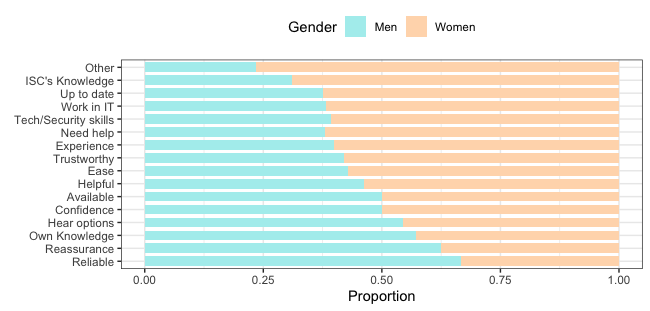} 
	\caption{Proportion of men versus women across rationales for YES to approaching ISC} 
	\label{fig:YES_by_gender}
\end{figure}

\begin{figure}[h]
	\centering
	\includegraphics[keepaspectratio, width=.9\columnwidth]{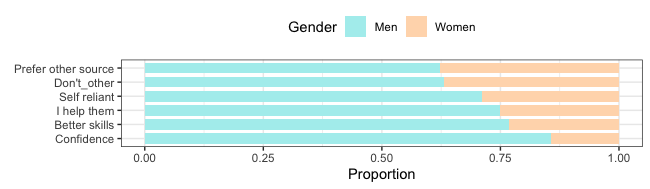} 
	\caption{Proportion of men versus women across rationales for NO to approach ISC} 
	\label{fig:NO_by_gender}
	\vspace{-.8cm}
\end{figure}

 
Of those who reported \emph{Yes} to approaching ISC, a higher proportion of women than men were motivated by the following perception of their ISC: 
general knowledge of ISC, ISC being up to date with latest technologies, ISC working in an IT related field, their technology / SP skills or experience as well as their trustworthiness, ease of access and helpfulness, as shown in Figure~\ref{fig:YES_by_gender}.
However, a higher proportion of men than women reasoned to approach ISC to hear various SP options, for reassurance, due to their own poor knowledge or perceived reliability of their ISC.
In comparison, Figure~\ref{fig:NO_by_gender} shows a higher proportion of men compared to women responding \emph{No} to approaching their ISC across all rationales 
(their confidence, better skills, self-reliance or preference for other sources, as well as being the one who helps others). 

\vspace{-.8cm}

\subsection{Advice Received from ISC}
\begin{OldVersion}
\begin{figure*}[h]
	\centering
	\includegraphics[keepaspectratio, width=.8\textwidth]{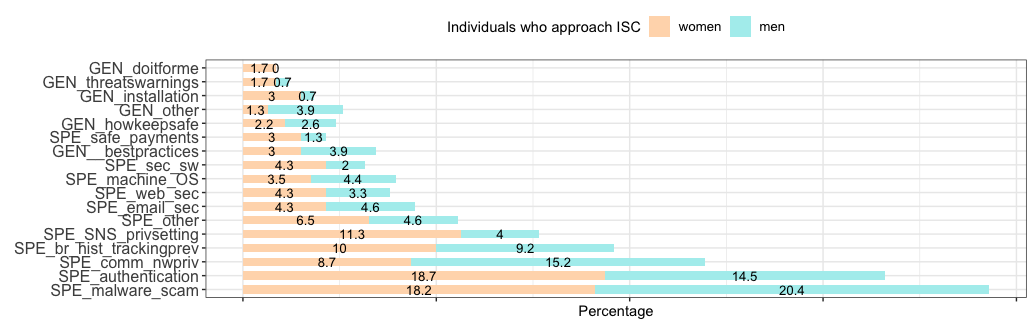} 
	\caption{\% of women and men across advice received from those who approach ISC}
	\label{fig:advice_received_bygender}
\end{figure*}
\end{OldVersion}
We investigate RQ5, that is look into the advice received by the $63\%$ (n=382) of participants who responded \emph{Yes} to approaching ISC for protective SP. 
As described in Section~\ref{sec:analysis}, advice were specific (such as about malware protection or communication privacy) or general (such as how to keep safe online).
We provide the list of specific advice, together with example responses, in Table~\ref{tab:advice_received_from_SoC} in the Appendix, and
summarise the advice received by women versus men in Fig~\ref{fig:advice_received_bygender}.
The main observable patterns 
are that (i) a higher \% of men than women receive advice about `\emph{malware / scam}' which included antivirus, virus threat; anti-malware, malware threat; anti-spyware, fraud and `communication and network privacy';
(ii) a higher \% of women receive advice about `\emph{authentication}', which included change passwords, use multiple passwords, use / set strong passwords, password manager or use MFA; and `privacy settings and SNS'.

\begin{RedundantContent}
\subsection{Additional Stuff}
\textbf{\textcolor{blue}{[LOOK AT THE QUESTIONNAIRE FOR ANY OTHER INPUTS THAT MAY HELP THE EQUALITY ANGLE OR THE PROTECTION-OUTCOME ANGLE... + Analyse the qual advice given]}}
- Q38: Confidence in ability (SE) to protect S\&P: men have a higher mean confidence - significantly different by t-test + can do a barchart comparison across gender across the 4 items; \textcolor{purple}{what does this mean: potentially that it links to advice source preference and use of tech/non tech SP? - a 1-pt increase in confidence predicts 12\% more likely to use techSP...
We find that those who say YES to approach SoC have a lower confidence, significant by t-test}\\
- Q41: how do you perceive the risks to your SP online (from 1-very low to 3-neither low or high and 5-very high)? - more women in the 3, slightly more men in the 4/5 \\
- Q42: open-ended: what are the risks to your SP online?\\
- men have a significantly higher average on risky behaviour scale\\
- Q44: more women (n=81) have installed an antivirus / antispyware compared to men (n=38)\\
- Q45: how often have updated an antivirus software -- more women on never/rarely/sometimes + more men on usually /many times within the past month --- this potentially may just say that women use auto-updates\\
- Q46: how often do you change passwords for financial and email accounts: more men on sometimes/usually + more women on never/rarely\\
- Q62: on SoC being accessible, having the skills, trusting SoC, on effectiveness and usefulness of SoC advice
\end{RedundantContent}

\section{Discussion}
\label{sec:discussion}
We summarise our findings and discuss how they extend existing literature, and their implications for equitable online safety. 

\noindent\textbf{Take-aways.}
The findings in broad terms are:
\noindent (1) there is an SP gender gap depicted by women and men's distinct reports of SP access and technology use patterns; 
\noindent (2) online advice is more associated with and predicts SP technology use than advice from ISC;
\noindent (3) women and men report distinct rationale for approaching ISC
and receive different SP advice.

\begin{Rebuttal}
(R#470D): We will name the broad take-aways which are (1) SP gap depicted by distinct SP access and technology use patterns; (2) online advice more associated with SP tech use than advice from ISC; (3) distinct rationale for approaching ISC. We are considering the fine-tunings around qualitative reporting proposed by (R#470D, R#470E).
\end{Rebuttal}

\begin{OldVersion}
\textbf{\emph{Findings Summary.}}
We noted key differences in the dynamics of SP protection across gender, namely:
\noindent (1) that women more often opt for SP protection advice from ISC, while men more often use online content and a combination of sources; 
\noindent (2) women report different engagement with SP technologies to men, who are more likely to use SP technologies; 
\noindent (3) those using online content are 3 to 11 times more likely to use SP technologies, as does an increase in the number of sources; 
\noindent (4) women and men differ in their motivation to approach ISC for SP protection advice;
\noindent (5) SP protection advice received differ between gender.
\end{OldVersion}


The social role theory explains that diverging social roles give rise to gendered expectations and stereotypes such as communal/supportive traits for women and agency/assertion for men, that in turn lead to different skills learnt and ways to behave~\cite{eagly2013sex}.
Our findings can be argued to emerge from gender stereotypes (whether personally endorsed or fitting into commonly held stereotypes), in particular
men less likely to approach ISC 
can be explained by masculine attitudes that discourage seeking help
~\cite{juvrud2017don} 
and SP stereotypes that men are less likely to ask for help and 
are over-confident (or women believed to be incompetent in SP and more comfortable to ask for help)~\cite{wei2023skilled}. 
Further, men more likely to report using online advice and engage with SP technology, relates to 
gender stereotypes in STEM 
where technology engagement is seen as a \emph{gender-type activity}~\cite{master2021gender,riegle2017shifting}, and further supported by assumptions of men being more interested and skilled in protection overall or the belief that SP is too complex for women who then delegate it to others~\cite{wei2023skilled}.
This view is supported by the fact that our male participants had higher affinity for new technical systems, compared to the women participants (as reported in Section~\ref{sec:participants}).
The affinity difference is likely a societal reflection,
but points to women being less inclined to self-explore (technological) SP.

In addition to their gender, people have racial, cultural, sexual and socio-economic identities that intersect and overlap with their gender identity~\cite{collins2020intersectionality,crenshaw1990mapping}. These factors 
confer a certain privilege or disadvantage for SP access, engagement and online safety outcomes (note related works~\cite{barwulor2021disadvantaged,walker2020more}), as well as impact perceptions and attitudes to SP~\cite{kwasny2008privacy} and resulting SP experiences (and needs)~\cite{geeng2022like}. 
While we did not control for these factors, 
education level was balanced between gender. The stark SP gender gap observed therefore demonstrates the relative importance of gender identity for online SP access and participation within user populations similar to our sample.
Further work is however needed to unpack 
the potential effects of how intersecting identities in the context of SP. 

\noindent \textbf{\emph{Implications for equitable online safety.}}
We add to evidence of a gender gap and un-equal SP experiences between women and men~\cite{oomen2008privacy,park2015men,coopamootoo2022feel}, and 
add a gender analysis to SP advice literature~\cite{rader2015identifying,rader2012stories,redmiles2017digital,redmiles2016learned}. 
This gap in protection already demonstrates an online safety divide that questions \emph{the equity of online safety opportunities}, that is whether SP access via online advice and SP technology affordances are appropriately configured for and serving women. 
Questions about the appropriateness of SP are supported by `gender blindness' arguments, 
where technology transformations are influenced by societal norms~\cite{barbieri2020gender} and the design and meaning of these technologies are created within gender relations and thus reflect pre-existing gender inequalities
~\cite{perez2019invisible}, such as only $17\%$ of technology jobs 
in Europe are held by women~\cite{europe2022gender}.
The large amount and varied type of SP advice online, that is known to overwhelm and lack prioritisation~\cite{redmiles2020comprehensive}, may also present more of a barrier for women's online safety than for men.

Although women in developed countries, such as the U.K, have equal opportunities for technology access and a high level of education~\cite{advanceHE2021equality,UKParliament2021equality,bbc2017record}, and 
the digital divide with respect to gender is said to be decreasing in developed countries~\cite{van2020digital}, this does not translate to equal online safety outcomes~\cite{ofcom2022report} or equal access and engagement with SP technologies as seen in this research.
The SP gap can be thought to be even worse in countries with wider digital divide.
Overall, this research supports reports of women being more at risk online~\cite{ofcom2022report} if the access and non-technology strategies that women employ do not result in equal online safety outcomes as men.

\begin{Rebuttal}
PART of Review Requirement 9: please include a discussion around intersectionality and how this may impact the results and recommendations of the work.
Rebuttal: (R#470D): We will discuss the potential influence of intersecting identities in the limitations and expand the sentence already in the discussion Section.
\end{Rebuttal}

\section{Actionable Recommendations}
We provide recommendations for stakeholders with an interest in the online safety ecosystem.

\begin{OldVersion}
that require (1) a concerted effort and dedicated resources contributing to 
collaboration between researchers across disciplines, educators, online advice providers (such as online safety advocates, governments, tech companies), tech industry, organisations supporting women communities, and policy makers, and 
(2) engaging with local and international bodies' who having been calling for action for a safe and equitable digital future and to keep women safe online~\cite{ofcom2022urgetechfirms,UNDP2022}. 
\end{OldVersion}


\noindent \textbf{Accessible \& effective online advice.} 
Our findings lead to questioning the relevance, accessibility and appropriateness of online safety advice, in particular those pointing to SP technologies as means of protection, for women.

\emph{First} we recommend efforts towards ensuring that the online advice ecosystem is inclusive of the various needs of the wide population of women, 
in addition to those tailored for specific threat scenarios such as intimate partner violence. 
The design of online safety advice need to be relevant to diverse women's assessment and response to threats.
Trustworthiness (as reported by our participants) and a sense of emotional support 
need to be designed within SP and digital advice affordances given (a) the affective dimension of SP~\cite{farahmand2019privacy,coopamootoo2017whyprivacy,stark2016emotional}, 
(b) women's higher likelihood for emotive evaluation of online threat scenarios~\cite{coopamootoo2022feel}, and (c) how the associated response actions provide a form of emotional coping~\cite{cho2020privacy,park2021users,jung2018investigation}.
This is supported by recommendations made in prior work~\cite{geeng2022like} that communication preceded by emotional support are of higher quality~\cite{feng2009testing}.
As practical example, the language used in online safety advice needs to be representative of the women groups it intends to serve, as opposed to being (overly) technical as previously reported~\cite{redmiles2020comprehensive}.

\emph{Second}, based on a complement of our findings and that of previous research raising issues of prioritisation and actionability of online SP advice~\cite{redmiles2020comprehensive}, 
as well as fragmentation across sources~\cite{redmiles2017digital}, we recommend standardising and continued revision of a key set of online sources and priority advice, given the current threat landscape affecting women.
Overall, the current lack of evidence on the effectiveness of online SP advice ecosystem for women (and diverse genders) needs to be addressed. 

\begin{MyNotes}
followup, trusting features
is it because women feel more assurance, emotional support with ISC? Geeng says 
Communication research has pointed out that advice pre- ceded by emotional support was considered higher quality [44].

\paragraph{Prioritise advice for women, given threat landscape}
Refine by ensuring they are: 
- inclusive in terms of applying to the general population + targeted, tailored advice for particular at risk group + personalised and be able to filter through what is relevant for particular community + threat model + threat scenario
- but now within the general itself we have a gender appropriateness issue
- hence online advice providers (advocates, governments, tech companies) need methods for ascertaining inclusivity in appropriateness and inclusivity
- apart from appropriateness, is the online advice relevant to women?

Women may have a more emotional evaluation of threat scenarios~\cite{coopamootoo2022feel} and needing emotional support to be associated with advice [geeng says creating affordances for collaborative discus- sion and feedback on advice documents may create better buy-in, sense of emotional support, and ability to archive out-of-date advice.]

What is the current top priority in online safety advice for women given active / predominant threat scenarios? most damaging / impactful threat scenarios?

- revise priorities + propagate new priorities given new threats 

- while there are women specific advice (see Wei2023 CHI) - we cannot yet measure the effectiveness of this 

\paragraph{Unify / make advice across sources consistent}
online, tv, ISC?

Geeng says we need specificity rather than consistency across sources as Reeder2017 `152 simple steps to stay safe online: Security advice for non-tech-savvy users' says

\paragraph{About fragmentation and multiple sources}
Men used more advice sources + had higher ATI

does the current advice ecosystem lead to multi-source checking which is helped by having a good ATI?

do stats on impact of ATI on advice source number
\end{MyNotes}

\textbf{Skills for SP.}
\begin{OldVersion}
\textcolor{purple}{put this comparison with the qual reporting to enrich the reports there}
\noindent \textbf{\emph{(1) Education, Skills \& the Digital Divide}}:
In addition, 
\end{OldVersion}
Compared to literature on the influence of \emph{technical web skills}~\cite{hargittai2012succinct} 
on advice source preference~\cite{redmiles2016learned}, our gender groups had similar digital skills across scales for information navigation, social and mobile skills, and differed slightly on operational and creative skills (measured via the internet skills scale~\cite{van2014measuring}), which raises questions about the digital skillset required for online safety.

\emph{First}, we recommend assessment and marking of the type and level of digital skill level needed to comprehend and action online safety and SP tech advice, including (in parallel) ways to develop these skills, such that people are supported rather than left to their own means of filling skill gaps.
This is linked to developing confidence in SP protection, given we noted women's poorer affinity for technical systems. 
\emph{Second}, for equity of SP opportunities, 
we recommend designing advice and SP technology engagement such that 
anyone can gain optimal protection irrespective of their skill level. 

\begin{MyNotes}
--- are online advice too filled with technical jargon, such that it is not clear how to action them?

Would be nice to have an assessment of what quality of skills is needed and have ways to inform/educate people about these in parallel so it's not left on the internet user to figure out how to understand the advice. For instance it could have advice targeting minimal skills to more advanced, where even the minimal would give paths to optimal protection.
\end{MyNotes}



\textbf{Socially supported SP.}
ISC advice, as we evidence, provides a valuable alternative to traditional individualistic SP design. 
Compared to burdening individual users with problem solving, it provides a collaborative, communal and supported version, which is particularly useful for coping with SP complexity or the (psychological and emotional) aftermath of attacks, including new ones where users may not yet have protection experience.
We make recommendations supporting the social SP body of research~\cite{wu2022sok,slupska2021participatory}.

\emph{First} we recommend SP technologies to have a socially-supported version and online safety advice to offer these given user preference.
\emph{Second} we recommend tech features that make it easy for anyone to ask for help or compare notes / options, whether from a known contact or anonymously.
We recommend defining a checklist of what to ask and for determining who to ask for SP advice, as 
the heuristics our women participants provided 
do not guarantee that ISC have SP skills. 
\emph{Third} we recommend learning from women's strategies
and the development of methods to sustain in-person dialogue and render it effective in supporting learning (such as bite-size template, multi-modal ways of delivery, protection evaluation and feedback - within online / offline spaces) 
and a re-envisioning of online SP that taps into and includes the SP patterns adopted by women, 
as well as for designers to consider the stereotypical cues in technology that cause gender-type digital engagement~\cite{cheryan2009ambient,murphy2007signaling}.


\begin{OldVersion}
\paragraph{Encourage any gender to ask for help}
Men do not ask for help in person - could they be overconfident?
--- perhaps they ask for help and compare notes online (in forums?)
How well did the \% who did not ask for ISC help do in engagement with SP?

While advice from ISC is not necessarily risky advice, most of the reasons provided by our women group for opting for ISC advice, such as work in IT, trustworthiness, or ease of access (complementing previous work on source perception~\cite{redmiles2016think}), do not guarantee ISC actually have SP skills to advise and enable learning (complementing previous work that family are not necessarily SP experts~\cite{murthy2021individually}) and may show an over-reliance, while the men group's rationale, such as evaluating options and reassurance, indicate a more analytical dialogue that can 
improve skills. 


SP advice from non-expert ISC runs the risk of (1) disseminating misinformation, similar to any topic where people provide information to each other~\cite{buchanan2020people}, 
(2) hampering agency by imposing the ISC's threat / protection model on the recipient~\cite{murthy2021individually},  
impeding skills development due to over-reliance and can be experienced as paternalistic and disempowering~\cite{murthy2021individually}, 
and (3) hindering the widespread use of certain SP technologies, given advice providers are not supported to share complex SP use \textcolor{purple}{[...teach complex SP interaction dynamics...]}, where SP technologies are usually designed for individual use and do not by default come with collaborative SP features, with a few recent research efforts in that direction~\cite{chouhan2019co,lipford2012someone} or the sharing of resource access~\cite{jacobs2016caring,watson2020we,moju2022you}. 
In contrast, while online content offers a breadth of quality~\cite{redmiles2020comprehensive}, our results question its accessibility for women.

\paragraph{Learning from women's strategies for online safety.}
e.g. mirror in-person (Geeng) / ISC dialogue, followup, trusting features
is it because women feel more assurance, emotional support with ISC? Geeng says 
Communication research has pointed out that advice pre- ceded by emotional support was considered higher quality [44].

Slupska et al. writes, ``Cybersecurity is more effective when it is communal...Discuss[ing] online threats and mitigations with members of a community makes it eas- ier and less intimidating to take action"
\end{OldVersion}

\textbf{Gender agenda.} 
We are far from addressing the gender gap and many unknowns remain. 
We strongly recommend a multi-perspective research agenda 
focused on understanding the role gender norms, stereotypes and intersecting identities play on SP opportunities, access and outcomes, and the role of gender theory within SP, akin~\cite{frener2022theorizing}. 
This requires (1) a concerted effort and dedicated resources, 
(2) collaboration between multi-disciplinary researchers and stakeholders such as policy makers, community support and SP advocacy groups, and 
(3) engaging with local and international bodies' who having been calling for action for a safe and equitable digital future and to keep women safe online~\cite{ofcom2022urgetechfirms,UNDP2022}. 

\textbf{Critical reflection on equity.}
Although the digital divide is thought to be growing weaker within developed countries 
and there are increased possibilities for SP access through online advice, useful and meaningful engagement with these to sustain SP usage and safety outcomes is yet to be addressed, in particular for women as this research shows, and  
the lack of prioritisation~\cite{redmiles2020comprehensive} and consensus about good advice~\cite{reeder2017152}. 


Given our findings point to a gender divide in SP opportunities and participation, demonstrated via gaps in access and use of SP technologies, we strongly recommend critical reflective action for the wider SP community of researchers, developers, technology providers, online safety advocates and policy makers, to address the question `\emph{for whom are we producing SP technology for?}'
With online safety considered a social good and its equity advocated by international human rights organisations~\cite{UNDP2022}, 
`\emph{what does gender equity in online safety involve in terms of SP opportunities, access, participation or outcomes?}'
Given women's personal, social and economic realities and their socio-cultural roles compared to men, `\emph{does equality of SP, such as even access methods, distribution and design, provide for equitable online safety outcomes?}' 

\textbf{Assurance of equity.}
Since gendered SP access and participation could disadvantage large strata of society,
we recommend the development of an \emph{SP equity assurance framework} and complementary tech policy,
that requires for instance (1) that online SP providers demonstrate assurance of equity in development and distribution, such as
consideration of the personal, social, and economic realities of women and their online safety needs; and 
(2) that SP advocates, policy makers, and researchers co-create equity markers and criteria.


\begin{OldVersion}
\textbf{\emph{(6) The Digital Divide}}:
Although the digital divide is thought to be growing weaker within developed countries (such as the U.K.)~\cite{van2020digital} and
there are increased possibilities for SP access through online advice, useful and meaningful engagement with these to sustain SP usage and safety outcomes is yet to be addressed, in particular by women as this research shows, and  
the lack of prioritisation~\cite{redmiles2020comprehensive} and consensus among experts on what is a good advice~\cite{reeder2017152}.

In consequence, a future weakening of the SP, protection and safety gaps online, would require addressing equity in a complex interconnection of factors, where first currently particular digital skillset are required to protect from harm online or for privacy and security decisions, such that although women in developed countries may have equal opportunities for access and a high level of education~\cite{advanceHE2021equality,UKParliament2021equality,bbc2017record}, this does not translate to equal protection outcomes~\cite{ofcom2022report}, and where the gap can be thought to be even worse in countries with wider digital divide. 
Second, perceptions and attitudes to SP~\cite{kwasny2008privacy} and resulting SP experiences and needs~\cite{geeng2022like} are likely different across gender, and reinforced by gender stereotypes in perceptions of SP~\cite{wei2023skilled} and national culture, which acts as a collective mindset that influence the perceptions and behaviours of individuals within nations~\cite{hofstede1984culture}. 
The issue of differential SP perception and experience between men and women, are further supported by `gender blindness' arguments, 
where societal norms influence technology transformations~\cite{barbieri2020gender} and where the design and meaning of these technologies are created within gender relations and thus reflect pre-existing gender inequalities
~\cite{perez2019invisible}, in particular where only $17\%$ of technology jobs, such as programming, systems analysis, or software development, in Europe are held by women~\cite{europe2022gender}.
\end{OldVersion}

\section{Conclusion}
This empirical research provides a gender analysis of online safety with a lens across protective SP access and protection outcome,
in particular showing that women's distinct SP access preference through ISC means that they cannot enjoy the breadth and quality of \emph{online} SP advice, that consequently lead to less sophisticated SP technology engagement, with potential impact on protection outcome (thereby demonstrating a gender SP discrepancy) and
adding to the few previous evidence of an SP gender gap. 
In consequence, this research (1) supports arguments of women being more at risk online compared to men, 
(2) supports local and international calls for action to keep women safe online, 
and (3) make recommendations for multi-stakeholder actions to ensure their protection.


{\footnotesize 
\bibliographystyle{plain}
\bibliography{repository,privacy,social-privacy,inequality,usermethods,repository2,emotion}
}

\appendix
\section{Education across gender}
\begin{table} [H]
\centering
\caption{\% Education level across gender} 
\label{tab_gender-education}
\footnotesize
\begin{tabular}{l|rr}
\toprule
\textbf{Educ. Level} & \textbf{\% Women} &  \textbf{\% Men}\\ 
\midrule
School incomplete &5.9 &4.0  \\
High School &24.4 & 25.2 \\
College &18.5&18.9 \\
Undergraduate & 34.7&41.9\\
Masters &14.5&9.0\\
PhD &2.0 & 1.0\\

\bottomrule
\end{tabular}
\end{table}

\section{Advice Source as Reported}
\emph{The blue row denotes higher \# of men, while the rose denotes higher \# of women.}
\definecolor{mistyrose}{rgb}{1.0, 0.89, 0.88}
\begin{table} [H]
\centering
\caption{Advice source for overall sample (N=604), with almost equal \# of women and men.} 
\label{tab:sources_by_gender}
\footnotesize
\begin{tabular}{l|rr}
\toprule
\textbf{Sources (n reports)} & \textbf{\% Women (n)} &  \textbf{\% Men (n)}\\ 
\midrule
\rowcolor{LightCyan} General research (163) & 41.1 (67)	& 58.9 (97) \\ 
\rowcolor{mistyrose} Family (101)&76.2 (77)& 23.8 (24)\\
\rowcolor{mistyrose}Friends  (97)& 50.5	(49)&49.5	(48) \\
\rowcolor{LightCyan} Specialist pages (67)& 23.9 (16)&76.1 (51)\\
\rowcolor{LightCyan} News, TV shows (65)& 44.6	(29)	&55.4	(36)\\
\rowcolor{LightCyan} Tech adverts/brand/mag (57)& 35.1	(20)&	64.9	(37) \\ 
\rowcolor{LightCyan} Online reviews (54)& 46.3	(25)&	53.7	(29)\\
\rowcolor{mistyrose} From work / colleagues (49)&65.3	(32)&	34.7	(17) \\
\rowcolor{LightCyan} Shared online content (48)& 33.3	(16)&	66.7	(32)\\ 
\rowcolor{LightCyan} Online forums (46) & 21.7	(10)&	78.3	(36) \\ 
\rowcolor{mistyrose} Face-to-face / offline  (34) &61.8	(21)&	38.2	(13) \\
\rowcolor{mistyrose} Training  (30)&  53.3	(16)&	46.7	(14) \\
\rowcolor{LightCyan} System prompts and settings (19) &47.4	(9)&	52.6	(10)  \\  
\bottomrule
\end{tabular}
\end{table}

\begin{Rebuttal}
\begin{table} [H]
\centering
\caption{Advice source for sub-sample with computing / IT background (n=100), with \emph{n=29} women and \emph{n=71} men.}
\label{tab:sources_by_gender-100}
\footnotesize
\begin{tabular}{l|rr}
\toprule
\textbf{Sources (n reports)} & \textbf{Women (n)} &  \textbf{Men (n)}\\ 
\midrule
\rowcolor{LightCyan} General research (35) & 9	&26 \\ 
\rowcolor{LightCyan} News, TV shows (18)& 7&	11\\
\rowcolor{LightCyan} Specialist pages (17)& 4	&13\\
\rowcolor{LightCyan} Online forums (15) & 1	&14 \\ 
\rowcolor{LightCyan} Training  (15)& 3	&12\\
\rowcolor{LightCyan} From work / colleagues (14)&4	&10 \\
\rowcolor{LightCyan} Shared online content (13)& 4	&9\\ 
\rowcolor{LightCyan}Online reviews (13)& 4&	9\\
\rowcolor{LightCyan} Tech adverts/brand/mag (12)&3	&9 \\ 
\rowcolor{LightCyan} Friends  (10)& 4	&6 \\
\rowcolor{mistyrose} Family (9)&6	&3\\
\rowcolor{LightCyan} Face-to-face / offline  (3) &1	&2 \\
\rowcolor{LightCyan} System prompts and settings (2) &0	&2 \\  
\bottomrule
\end{tabular}
\end{table}
This table should be re-interpreted, as we don't have equal number of women and men
\end{Rebuttal}

\section{SP technologies and methods usage}
\label{sec:appen_SP_list}

\begin{table}[H]
\centering
\caption{Security \& Privacy Technologies/Methods}
\label{tab:SP_categories}
\footnotesize
\resizebox{\columnwidth}{!}{
\begin{tabular}{ll|r}
\toprule
\textbf{Category} &\textbf{SP Methods} & \textbf{\% of N=604} \\
\midrule
&Antivirus& 40.7 \\ 
   & Passwords&56.2\\
  &  Firewall&16.7\\
   & VPN/Proxy&12.6\\
  &  Anti-tracking&11.1\\
   & Antimalware&10.9\\
  &  Private browsing/Incognito&11.4\\
Technological&    Auth (2FA, MFA)   &9.3\\ 
  &  Privacy/security settings&8.9\\
  &  Security suite/security brands&5.3\\
  &  Cookies&4.0\\
  &  Security and software updates&2.1\\
  &  HTTPS (filtering)/site security&1.9\\
 &   Antispyware     &1.7 \\
    &Encryption     &1.4\\
  &  Tor&0.5\\ 
  &Built-in security&6.1\\ 
  &   Other tech  &2.1\\ \midrule	


 &   Limit online engagement&5.8\\
Non-technological&    Not open/click random emails and links &4.3\\
  &  Different screen names  &3.6\\
 &    ‘Common sense'&1.8\\
 &   Reporting phishing emails&0.7\\
  &   Other non-tech  &2.5\\ \midrule
 None / & Nothing/‘I do not know'&16.0\\
   other & ‘Something but do not remember'&1.7\\
\bottomrule
\end{tabular}
}
\end{table}
We note that `other tech' refers to responses naming use of PayPal, notifications,    cloud storage or backup systems, while `other non-tech' refers to not storing information online, locking screen, using only one device, logging out of account, avoiding suspicious sites and using only secure connections.
\begin{table} 
\centering
\caption{\% women vs men reporting to use at least 1 SP method (\emph{overall N=604; women n=303; men n=301})}
\label{tab:atleast1SP-by-gender}
\footnotesize
\begin{tabular}{l|rrr}
\toprule
\textbf{SP method type} &\textbf{\% Overall} &  \textbf{\% Women} &  \textbf{\% Men}\\ 
\midrule
Tech SP &80.8 & 74.3 &87.4  \\
Non-tech SP & 16.1 & 17.5& 14.6 \\
Nothing&17.5 & 22.8&12.3 \\
\bottomrule
\end{tabular}
\vspace{-5mm}
\end{table}

\section{Depiction of Quasi-separation}
\begin{figure}[H]
	\centering
	\includegraphics[keepaspectratio, width=.7\columnwidth]{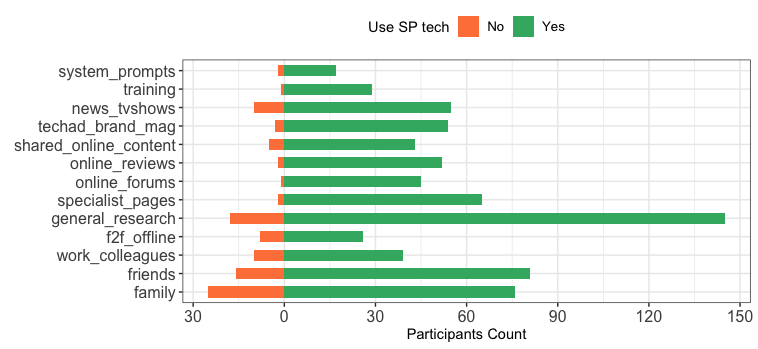} 
	\caption{Depicting data separation in use of SP technologies by advice source preference}
	\label{fig:quasi_data_separation}
\end{figure}

\section{Example Responses \& Codebook}
\onecolumn
\begin{table}
\centering
\caption{Example responses for Advice Sources named}
\label{tab:e.g.response_advice}
\footnotesize
\resizebox{.8\textwidth}{!}{
\begin{tabular}{ll}
\toprule
\textbf{Advice Source} & \textbf{Example participant response} \\ \midrule
\rowcolor{apricot}Family & ``\emph{through my \textbf{husband}, it is his job}" (W38)\\ 
\midrule
\rowcolor{apricot}Friends & ``\emph{recommendations from \textbf{friends} who are more tech savvy}" (M18) \\
\midrule\rowcolor{apricot}Face-to-face / offline &  ``\emph{i only hear of them from \textbf{the man who fixes my laptop}}" (W23) \\
\midrule\rowcolor{apricot}From work & ``\emph{from my \textbf{employers} IT department}" (M9)  \\
\midrule\rowcolor{apricot}Colleagues & ``\emph{I read articles shared by \textbf{colleagues} that work in cyber security,} \\
\rowcolor{apricot}& \emph{and make sure i follow the same security companies they do}" (W65) \\

\midrule
\rowcolor{lightblue}General research & (a) ``\emph{I usually find out by either googling them online myself...}" (M46) \\
\rowcolor{lightblue}&(b) ``\emph{I search the internet for the latest most reliable systems}" (M170)\\
\midrule
\rowcolor{lightblue}Specialist pages e.g. Techcrunch & (a) ``\emph{IT magazine sites \& blogs}" (M10)\\
\rowcolor{lightblue}& (b) ``\emph{I learn from Tech Crunch, Buzzfeed and general news media reportage ...}" (M48)\\
\midrule
\rowcolor{lightblue}Online reviews and recommedations & ``\emph{I check reviews for online protection and stay with same company if all has gone well in the past}" (M54)\\
\midrule
\rowcolor{lightblue}Tech adverts, shared company info, & (a) ``\emph{I usually see adverts online about protecting privacy...}" (M132)\\ 
\rowcolor{lightblue}reputable brands & (b)``\emph{...Also get safe browsing info and vpn info sent to my inbox from email companies I use such as ProtonMail}" (M60) \\ 
\rowcolor{lightblue}&(c) ``\emph{from microsoft [sic] and other companies themselves}" (W160), ``\emph{I tend to use well known brands such as Norton}" (M172)\\ 
\midrule
\rowcolor{lightblue}Social media content & (a)``\emph{... eg on linked in [sic], they share security related articles and i check this at least weekly}" (W65)\\ 
\rowcolor{lightblue}e.g. YouTube, Twitter &(b)``\emph{I usually hear about them via twitter networks...}" (M229)\\
\rowcolor{lightblue}&(c)``\emph{...the Police and other agencies post information on social media which I follow}" (W362)\\ 
\rowcolor{lightblue}&(d)``\emph{I am subscribed to tech youtubers, who ways to protect}"(M449)\\ 

\midrule
\rowcolor{lightblue}Online forums & (a) ``\emph{...Various online communities and forums dedicated to technology and computer use}" (M578)\\
 \rowcolor{lightblue}&(b) ``\emph{I usually hear about it on reddit}" (W604) \\

\midrule
\rowcolor{lightkhaki}
News, TV shows & ``\emph{occasionally from articles I read in the newspaper / online}" (W57), ``\emph{by seeing articles about them online,} \\
\rowcolor{lightkhaki}& \emph{especially on the BBC website}" (W497), ``\emph{watching the News}" (M180),  ``\emph{usually via reliable sites such as BBC or Which}\\
\rowcolor{lightkhaki}&  \emph{or consumer rights programmes (Martin Lewis etc)}" (W16)\\


\midrule
\rowcolor{lightkhaki}
Training  & (a) ``\emph{... an A level course in computers systems and networking}" (M394) \\
\rowcolor{lightkhaki}&(b) ``\emph{I have been bought up learning about the need for protection information online}" (M405)\\ 
\rowcolor{lightkhaki}&(c) ``\emph{I research software engineering and cyber security, it's part of my job}" (M239)\\

\midrule
\rowcolor{lightkhaki}System prompts and settings & (a) ``\emph{I own apple devices and I rely on their regular updates for updating privacy protection on my electronic devices. I don't }\\
\rowcolor{lightkhaki} &\emph{read details of what protection this provides}" (W39) \\  

\rowcolor{lightkhaki} & (b) ``\emph{I use the settings and FAQ in specific apps or software}" (M439), ``\emph{...Other times I will check out privacy/security}" \\
\rowcolor{lightkhaki} &settings to see what I need to do" (W412)\\

\midrule
\rowcolor{lightkhaki}
Consumer magazine & ``\emph{Which? magazine is a good source}" (W24)  \\ 

\bottomrule
\end{tabular}
}
\end{table}

\begin{table}[H]
\centering
\caption{Rationale for ISC advice with example participant responses}
\label{tab:e.g.response_approach_ISC}
\footnotesize
\resizebox{.85\textwidth}{!}{
\begin{tabular}{llll}
\toprule
&\textbf{Category} &\textbf{Reason} & \textbf{Example participant response} \\ 
\midrule
\rowcolor{celadon}Yes&ISC&(General) Knowledge & (a) ``\emph{Because they are usually very knowledgeable}" (W41)\\ 
\rowcolor{celadon}&Skills&of ISC& (b) ``\emph{Usually ask the kids as they know so much more than me}" (M114) \\
\rowcolor{celadon}&&& (c) ``\emph{My friends and family tend to know about these things before me so i go them for advice}" (M352)\\

\midrule
\rowcolor{celadon}&&Technology skills & (a) ``\emph{My daughters \textbf{know more about computers} so I ask them}" (W94)\\
\rowcolor{celadon}&&of ISC & (b) ``\emph{Because I live with my partner who is very \textbf{IT security aware} ...}" (W37) \\

\midrule
\rowcolor{celadon}&&ISC works in IT&(a) ``\emph{Husband \textbf{works in IT}}" (W28), 
``\emph{My husband helps me with all this.  He is an IT engineer}" (W470)\\ 
\rowcolor{celadon}&&& (b)``\emph{I have \textbf{friends that are computer programmers} so I know they will tell me what I need to do}" (W29) \\
\rowcolor{celadon}&&& (c) ``\emph{Most of my family and friends and coworkers \textbf{all work in IT Security / Networking jobs}}" (W65)\\ 

\midrule
\rowcolor{celadon}&&Experience of ISC&(a) ``\emph{They have greater experience than me}" (M147)\\
\rowcolor{celadon}&&&(b) ``\emph{There's usually someone who's been in that situation before and who doesn't mind sharing what they did}" (W72) \\

\midrule
\rowcolor{celadon}&& Up to date& ``\emph{[they] are the most up to date and informed on the subject}" (M150)\\

\midrule
\rowcolor{teagreen}&ISC&Trustworthy& (a) ``\emph{Because a friend and family member are trustworthy connections}" (M137) \\
\rowcolor{teagreen}&Qualities&&(b) ``\emph{My father is very experienced with computers and I trust his opinion}" (W560) \\

\midrule
\rowcolor{teagreen}&&Easy to access& ``\emph{Because I find it easier to get advice from others}" (M112),
``\emph{they know quicker than me finding it}" (W191)\\

\midrule
\rowcolor{teagreen}&&Helpful & (a) ``\emph{They have helped me improve my online habits tremendously}" (M575)\\
\rowcolor{teagreen}&&&(b) ``\emph{As they generally offer good advice}" (W283)\\

\midrule
\rowcolor{teagreen}&&Available& (a) ``\emph{[he is a] computer techie fellah.  He's easily available}" (W230) \\
\rowcolor{teagreen}&&& (b) ``\emph{Because my friend is always contactable}" (W209) \\

\midrule
\rowcolor{teagreen}&&Reliable&``\emph{they are the most reliable}" (M12), ``\emph{I can rely on my family}" (M140)\\

\midrule
\rowcolor{celadon}&Other& Need help & ``\emph{I just need support sometimes}" (M231), ``\emph{If I’m worried I ask my son what I should do}" (W225) \\

\midrule \rowcolor{celadon}&Skills& Knowledge & ``\emph{because I'm not that tech savvy and learning about online technologies doesn't really interest me at all}" (M18)\\

\midrule \rowcolor{celadon}&&Confidence & ``\emph{Not confident enough to do on my own}" (F102)\\
\midrule
\rowcolor{teagreen}&Other&Hear options/mutual & (a) ``\emph{
to see what options are out there}" (M572), ``\emph{Because I can get a variety of options and advice from them}" (M302)\\
\rowcolor{teagreen}&&&(b) ``\emph{Me and my friends will always share useful information with each other about anything happening online}"(M152)\\

\midrule
\rowcolor{teagreen}&&Reassurance& (a) ``\emph{It's better to get a second opinion on things, since you might have missed some important detail}" (M186) \\
\rowcolor{teagreen}&&&(b) ``\emph{I like to get reassurance that what I'm doing is correct and it is the same or similar to how they are doing it}" (M46)\\
\rowcolor{teagreen}&&&(c) ``\emph{Because I want to be sure that I am doing the correct thing}" (W590)\\
\midrule
\rowcolor{peach-orange} No &Own& Self reliance &  (a) ``\emph{I can rely on my own research}" (M74), ``\emph{I can do it myself}" (M95)\\
\rowcolor{peach-orange}&Skills&&(b) ``\emph{I tend to be quite self sufficient and find things out on my own}" (M517),
``\emph{I can figure things out on my own}" (W555)\\

\midrule
\rowcolor{peach-orange} && Better skills & (a) ``\emph{I am the IT expert}" (M8), 
``\emph{I know far more about it than most friends, family, and coworkers}" (M32)\\ 
\rowcolor{peach-orange} &&& (b) ``\emph{because they know less than me}" (M2), ``\emph{I don't think my friends / family know much about anything privacy related}" (M131)\\

\midrule
\rowcolor{peach-orange}&& Confidence& ``\emph{I am confident in finding information myself}" (M107),
``\emph{I'm confident in my own ability to find what I need}" (M127)\\
\midrule
\rowcolor{peach-yellow}&Other&Prefer other source & ``\emph{Usually there is a wealth of expert information online so there is no need to consult family and friends}" (M35)\\
\rowcolor{peach-yellow}&&&``\emph{i prefer government websites for info}"(M1),  ``\emph{I would rather research it myself online from a variety of sources}" (M122)\\

\midrule
\rowcolor{peach-yellow}&&Do not need help / other & ``\emph{Because I don't feel the need too}" (M602),
``\emph{I don't often need help with privacy/security}" (M268),
``\emph{Never lost data}" (M490)\\

\midrule
\rowcolor{peach-yellow}&& The one helping others&``\emph{Because I am the one friends/family tend to go to}" (M178), ``\emph{They would expect me to be able to tell them}" (M10) \\
\bottomrule
\end{tabular}
}
\end{table}

\begin{table}[H]
\centering
\caption{`Specific' advice / support received by those participants who approach ISC (\% participants column are from those who approach ISC only)}
\label{tab:advice_received_from_SoC}
\footnotesize
\resizebox{\textwidth}{!}{
\begin{tabular}{lcl}
\toprule
\textbf{Advice Topic} & \textbf{\% participants} & \textbf{Example participant response} \\ \midrule
anti-virus, -malware and scam advice & 19.1 & 
``about the different ways one can be catfished" (M394) \\ 

\rowcolor{LightCyan}
authentication & 17.0 & ``... suitable password use" (M62); ``... I was recommended Lastpass by a family member" (M162) \\
\rowcolor{LightCyan}
&& ``... Advice to use a different password for each website I use" (W81)\\

communication, n/w privacy & 11.2& ``...setting up vpn for instance" (M15); ``router security setup" (M36); ``Use https..." (M64)\\
&& ``Help with VPNs and proxy" (M158); ``...which Firewalls are the best and how to update my security settings..." (W427)\\

\rowcolor{LightCyan}
browser history, tracking prevention & 11.2 & ``assisting me in setting up an ad blocker \& also reminding me how to get rid of cookies stored on my laptop" (W37)  \\
\rowcolor{LightCyan}
&& ``Clearing history and cookies" (M62); ``Installing adblocks etc" (W102); ``clear cache, incognito" (M138)\\
\rowcolor{LightCyan}
&& ``How to implement private browsing and control your cookies" (M142); ``I first learned about Ghostery via friends" (M268)\\
\rowcolor{LightCyan}
&& ``my dad mentioned ublock as a good option to block adverts" (W321)\\

privacy setting / SNS & 8.4 & ``how to set up privacy settings online" (W543); ``advise on accessing and changing privacy settings" (W501)\\
&& ``Facebook and twitter posts" (M528); ``setting privacy on facebook" (M493)   \\

\rowcolor{LightCyan}
other specific & 5.8 & ``My close friend is currently very concerned about voice assistants, e.g. Alexa, so all the microphones are shut off in\\
\rowcolor{LightCyan}
&& all their devices as much as possible" (W4); ``advice about opting out of communication related to marketing" (W39)\\
\rowcolor{LightCyan}
&& ``advice about how small pieces of personal info can be put together" (W69)\\
\rowcolor{LightCyan}
&& ``...how to minimise the risk of unintentionally putting personal information online" (M144)\\
\rowcolor{LightCyan}
&& ``about fraud cases and examples of stories where other people" (M150)\\

email security & 4.5 & ``They also told me about having multiple emails accounts" (M180); ``Using alias emails" (W226)\\
&& `` be wary of opening emails from unknown or suspicious sources" (W210); ``Using bogus emails" (M214) \\

\rowcolor{LightCyan}
web security & 3.5 & ``use familiar websites"; P396 ``check websites, not give too much away" (W382)\\

system / OS & 3.7 & ``How to securely delete a mobile operating system" (M10); ``My partner set up my computer and its protection" (M73)\\
&& ``Keep my laptop up to date on new updates and security issues" (W41)\\

\rowcolor{LightCyan}
security software & 3.4 &``which apps to use, how to use them and how to keep them up to date" (W108); ``Instructions on software and tips" (W282) \\
\rowcolor{LightCyan}
&& ``They advised certain softwares and gave me the pros and cons of different software packs" (W281)\\
\rowcolor{LightCyan}
&& ``My friend has helped me set up security for my online usage" (W195); ``Discussed apps they use" (M346)\\

safe payments & 2.3 & ``My family set up any new technology I need. e.g. pay pal" (M135) \\
&& ``about how to use your card safe online shopping" (M351); ``Idea of using PayPal instead of online banking" (W550)\\

\bottomrule
\end{tabular}
}
\end{table}
\begin{RedundantContent} 

 \begin{figure*}
    \centering
    \includegraphics[height = 5.5cm,width=.8\textwidth]{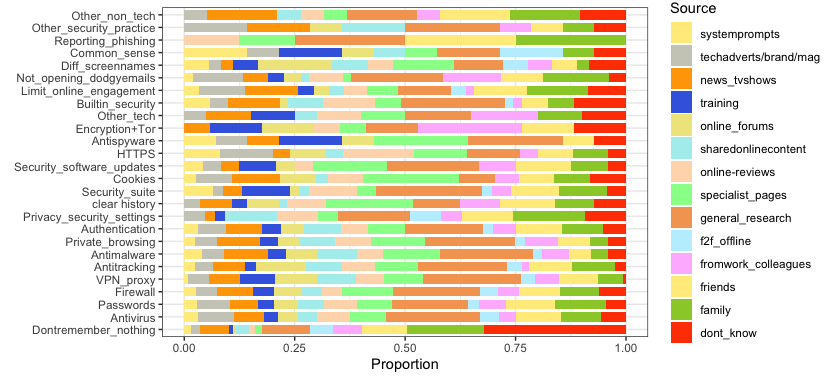}
    \caption{Proportion of Sources across each SP method \textcolor{purple}{[re-label Authentication to MFA throughout + n in each SP y-axis]}}
    \label{fig:SPMethodsxSources}
\end{figure*}

\end{RedundantContent}

\section{Example Advice Received by Gender}
\begin{figure*}[h]
	\centering
	\includegraphics[keepaspectratio, width=.8\textwidth]{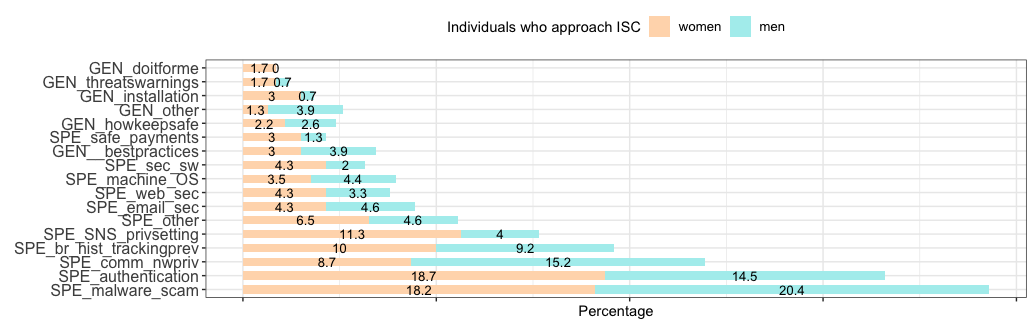} 
	\caption{\% of women and men across advice received from those who approach ISC}
	\label{fig:advice_received_bygender}
\end{figure*}

\twocolumn
\section{Questionnaire}

Part1: What privacy and security methods or tools do you most often use online? 
[If you are not aware or don't use any, please say so.]

Part2: How do you usually become aware of, learn about or find technologies/methods for protecting your privacy and security online?
[If you are not aware of any technologies/methods for protecting your privacy/security, please say so.]

Part3: At any time, if you need advice/help to protect your privacy and security online, do you usually approach your intimate and social connections (such as family, friends, coworkers) for help? 
Please explain why you replied Yes or No to the previous question.

Part4: If you have received advice/support from intimate and social connections, about technologies/methods to protect your privacy/security online, please provide examples of these advice/support.

Part5: (i) What is your gender? $\square$ Woman $\square$ Man $\square$ Non-binary $\square$ Other

(ii) What is your age?

(iii) What is your ethnicity? $\square$ White $\square$ Mixed $\square$ Asian: Indian $\square$ Asian: Chinese $\square$ Asian: Other $\square$ Black $\square$ Other ethnic group

(iv) What is the highest level of education that you have completed?
$\square$ Did not attend school
$\square$ Attended school but did not graduate
$\square$ Graduated from high school
$\square$ Graduated from college
$\square$ Undergraduate degree
$\square$ Masters degree
$\square$ PhD

(v) Do you have an education in, or work in, the field of IT, computer science or computer engineering?
$\square$ No
$\square$ Yes (please specify your level of education in computing/IT or your area of work.

(vi) Please respond to the following statements with [Not at all true of me] to [Very true of me].
$\square$ Not at all true of me
$\square$ Mostly true of me
$\square$ Neither true nor untrue of me
$\square$ Mostly true of me
$\square$ Very true of me\\
I know how to open downloaded files\\
I know how to download/save a photo I found online\\
I know how to use shortcut keys (e.g. CTRL-C for copy, CTRL-S for save)\\
I know how to open a new tab in my browser\\
I know how to bookmark a website\\
I know where to click to go to a different webpage\\
I know how to complete online forms\\
I know how to upload files\\
I know how to adjust privacy settings\\
I know how to connect to a WIFI network\\
I find it hard to decide what the best keywords are to use for online searches\\
I find it hard to find a website I visited before\\
I get tired when looking for information online\\
Sometimes I end up on websites without knowing how I got there\\
I find the way in which many websites are designed confusing\\
All the different website layouts make working with the internet difficult for me\\
I should take a course on finding information online\\
Sometimes I find it hard to verify information I have retrieved\\
I know which information I should and shouldn’t share online\\
I know when I should and shouldn’t share information online\\
I am careful to make my comments and behaviours appropriate to the situation I find myself in online\\
I know how to change who I share content with (e.g. friends, friends of friends or public)\\
I know how to remove friends from my contact lists\\
I feel comfortable deciding who to follow online (e.g. on services like Twitter)\\
I know how to create something new from existing online images, music or video\\
I know how to make basic changes to the content that others have produced\\
I know how to design a website\\
I know which different types of licences apply to online content\\
I would feel confident putting video content I have created online\\
I know which apps/software are safe to download\\
I am confident about writing a comment on a blog, website or forum\\
I would feel confident writing and commenting online\\
I know how to install apps on a mobile device\\
I know how to download apps to my mobile device\\
I know how to keep track of the costs of mobile app use

(vii) In this question, we will ask you about your interaction with technical systems. The term ‘technical systems’ refers to apps or other software applications, as well as entire digital devices (e.g. mobile phone, computer, TV, car navigation).

Please indicate the degree to which you agree/disagree with the following statements.
$\square$Completely disagree
$\square$Largely disagree
$\square$Slightly disagree
$\square$Slightly agree
$\square$Largely agree
$\square$Completely agree\\
I like to occupy myself in greater detail with technical systems.\\
I like testing the functions of new technical systems.\\
I predominantly deal with technical systems because I have to.\\
When I have a new technical system in front of me, I try it out intensively.\\
I enjoy spending time becoming acquainted with a new technical system.\\
It is enough for me that a technical system works; I don’t care how or why.\\
I try to understand how a technical system exactly works.\\
It is enough for me to know the basic functions of a technical system.\\
I try to make full use of the capabilities of a technical system.

\end{document}